\documentclass[%
 reprint,
 superscriptaddress,
 amsmath,amssymb,
 aps,
 prb,
 floatfix,
]{revtex4-1}

\usepackage{array}
\newcolumntype {s}[1]{@{\hspace*{#1}}}
\newcolumntype{L}{>{\displaystyle}l}
\newcolumntype{d}[1]{D{.}{.}{#1}}
\newcommand{\mc}[1]{\multicolumn{1}{c}{#1}}
\usepackage{dcolumn}
\usepackage{amsmath}
\usepackage{bm}
\usepackage{graphicx}
\usepackage{epstopdf}

\newcommand* {\ket}[1]{| {#1} \rangle}
\newcommand* {\braket}[1]{\langle {#1} \rangle}
\DeclareMathOperator{\cp}{cp}
\newcommand* {\frack}[2]{{\textstyle\frac{#1}{#2}}}

\begin{document}

\title{Spin-orbit interactions in inversion-asymmetric two-dimensional hole systems: A variational analysis}

\author{E. Marcellina}
\affiliation{School of Physics, The University of New South Wales, Sydney 2052, Australia}

\author{A. R. Hamilton}
\affiliation{School of Physics, The University of New South Wales, Sydney 2052, Australia}

\author{R. Winkler}
\affiliation{Department of Physics, Northern Illinois University, DeKalb, IL 60115, USA}
\affiliation{Materials Science Division, Argonne National Laboratory, Argonne, IL 60439, USA}

\author{Dimitrie Culcer}
\affiliation{School of Physics, The University of New South Wales, Sydney 2052, Australia}

\date{\today}

\begin{abstract}
  We present an in-depth study of the spin-orbit (SO) interactions
  occurring in inversion-asymmetric two-dimensional hole gases at
  semiconductor heterointerfaces. We focus on common semiconductors
  such as GaAs, InAs, InSb, Ge, and Si. We develop a semianalytical
  variational method to quantify SO interactions, accounting for
  both structure inversion asymmetry (SIA) and bulk inversion
  asymmetry (BIA). Under certain circumstances, using the
  Schrieffer-Wolff (SW) transformation, the dispersion of the ground
  state heavy hole subbands can be written as $E(k) = A k^2 - B k^4
  \pm C k^3$ where $A$, $B$, and $C$ are material- and
  structure-dependent coefficients. We provide a simple method of
  calculating the parameters $A$, $B$, and $C$, yet demonstrate that
  the simple SW approximation leading to a SIA (Rashba) spin
  splitting $\propto k^3$ frequently breaks down. We determine the
  parameter regimes at which this happens for the materials above
  and discuss a convenient semianalytical method to obtain the
  correct spin splitting, effective masses, Fermi level, and subband
  occupancy, together with their dependence on the charge density,
  and dopant type, for both inversion and accumulation layers. Our
  results are in good agreement with fully numerical calculations as
  well as with experimental findings. They suggest that a naive
  application of the simple cubic Rashba model is of limited use in either
  common heterostructures or quantum dots. Finally, we find
  that for the single heterojunctions studied here the magnitudes of
  BIA terms are always much smaller than those of SIA terms.
\end{abstract}

\pacs{Valid PACS appear here}
\maketitle

\section{Introduction}
\label{sec:Introduction}

The ability to harness the spin degree of freedom is essential for
the development of practical semiconductor spintronic devices
\cite{Datta1990, Zutic2004} and quantum information
processing. \cite{Loss1998, Bulaev2005, Awschalom2013}
All-electrical spin control may be possible by exploiting the
coupling of the spin and orbital degrees of freedom brought about by
the strong spin-orbit interactions in certain semiconductor
systems. \cite{Datta1990, Zutic2004, Winkler2003} This could lead to
faster spin rotations and lower power consumption, as well as the
convenience of using solely electric fields, which are easier to
apply and localize than magnetic fields. The search for
semiconductor systems with strong spin-orbit coupling has led
naturally to low-dimensional hole systems. \cite{Winkler2000, 
  Papadakis2000, Fertig2003} Despite the promising advances of
recent years, in particular in the experimental state of the art,
\cite{Rokhinson2004, Chen2010, Klochan2010, Keane2011, Komijani2013, 
  Li2013, Spruijtenburg2013, Srinivasan2013, Nichele2014, Tracy2014, 
  Li2015, Mueller2015, Brauns2016} functional hole spin-based
devices are yet to be realized. In particular, a comprehensive
understanding of the interaction between a hole's spin and its
solid-state environment is far from complete.

In group IV and III-V semiconductors the uppermost valence band is
described by wave functions originating from bonding atomic $p$
orbitals with orbital angular momentum $L = 1$.  Together with the
spin $S = 1/2$ this results in an effective spin $J = 3/2$,
\cite{Winkler2003, Yu1996, Luttinger1956} which brings about spin
properties of hole systems that are distinct from those of electron
systems. \cite{Culcer2006, Culcer2007} The bulk valence band
eigenstates are heavy holes (HH) with angular momentum projection
$m_J = \pm 3/2$ in the direction of the wave vector and light holes
(LH) with $m_J = \pm 1/2$. These are degenerate at the zone center
but split by a finite energy at nonzero wave vectors. Confinement to
the interface of a heterojunction fixes the hole spin quantization
axis in the growth direction $\hat{\bm z}$, which here we take to be
parallel to $(001)$, and lifts the degeneracy of the HH and LH
states at the subband edge (in-plane wave vector ${\bm k} = 0$). At
this point the eigenstates remain pure HH and LH, while at finite
${\bm k}$, the $\bm{k} \cdot \bm{p}$ interaction causes the HH and
LH states to mix. However, for typical Fermi wave vectors $k_F$, the
subband states with $k \lesssim k_F$ can still be regarded as
approximately HH- or LH-like. \cite{Winkler2008}

Inversion asymmetry further lifts the spin degeneracies of HH and LH
states, \cite{Winkler2003} resulting in additional
${\bm k}$-dependent energy level splittings. Inversion asymmetry in
semiconductor heterostructures may stem from the asymmetry of the
confining potential [structural inversion asymmetry (SIA) or Rashba
terms \cite{Rashba1984}] or from the asymmetry of the underlying
crystal structure [bulk inversion asymmetry (BIA) or Dresselhaus
terms \cite{Dresselhaus1955}]. BIA and SIA spin splittings are
proportional to odd powers of $k$, with $k$-linear\cite{Rashba1988, 
Manaselyan2009, Bi2013, Winkler2003, Luo2010, Durnev2014, Wenk2016}
and $k^3$ terms\cite{Dresselhaus1955, Winkler2000, Winkler2002, 
Winkler2003, Bulaev2005, Sugimoto2006, Schliemann2006, Winkler2008, 
Wong2010, Bi2013, Sacksteder2014, Biswas2015, Wenk2016} frequently
representing the dominant contributions. They result in a high
degree of nonparabolicity of the hole energy bands. Accounting for
the complex couplings between the hole bands is vital if one is to
capture all aspects of hole spin dynamics correctly. So far,
theoretical studies of two-dimensional hole gases (2DHGs) in group
IV and III-V structures have been predominantly numerical and always
material-specific. \cite{Rashba1988, Broido1985, Ekenberg1984, 
Volkov1998, Goldoni1995, Bangert1985, Winkler2002}
Calculations for single heterojunctions are more complex
due to the nature of the confining potential.  Unlike quantum wells,
where the potential shape is approximately fixed, the shape of the
confining potential in single heterojunctions is highly
density-dependent, thus requiring self-consistent wave functions
that produce the potential for a given density.

In this paper, we develop a variational method that enables us to
gain a transparent insight into spin-orbit interactions in 2DHGs in
III-V and Si-based heterojunctions. We use the Luttinger Hamiltonian
\cite{Luttinger1956} and the standard envelope function
approximation, \cite{Winkler2003} combined with a simple
self-consistent variational approach \cite{Fang1966, Kelly1990, 
  Winkler1992} to calculate the spin splitting and effective masses
for semiconductors with zincblende and diamond lattices. The
variational approach allows one to easily solve for the confinement
potential $V(z)$. Once the Luttinger Hamiltonian for the 2DHG is
constructed, one can use a Schrieffer-Wolff (SW) transformation
\cite{Schrieffer1966} (in the context of semiconductors also known
as L\"{o}wdin perturbation theory \cite{Lowdin1951}) to derive
analytical expressions for the spin-dependent dispersion of
2D hole systems, which subsequently yields the Fermi level, spin
splitting, effective masses and subband occupancy. We apply this
method to common semiconductors such as GaAs, InAs, InSb, Ge, and
Si, and compare inversion and accumulation layers.

In the axial approximation, i.e., ignoring anisotropic corrections
including warping and BIA, the SW transformation enables one to
write the dispersion relation in a rather simple form $E_\pm (k) = A
k^2 - B k^4 \pm C k^3 $ where $A$, $B$, and $C$ are material- and
structure-dependent coefficients, and the $C k^3$ term represents
the Rashba spin splitting. \cite{Winkler2000, Winkler2002, 
  Winkler2003, Bulaev2005, Schliemann2006, Sugimoto2006, 
  Winkler2008, Wong2010, Bi2013, Sacksteder2014, Biswas2015}
However, we demonstrate that the validity of the SW-transformed model is limited to a relatively
narrow range of parameters, indicating that in general the HH spin
splitting contains higher-order terms in the wave vector, which are frequently
sizable. The limited applicability of the simple dispersion
relation to realistic heterostructures is relevant to the current
understanding of the spin-Hall conductivity, \cite{Schliemann2004, 
  Sugimoto2006, Wong2010, Bi2013} hole spin helix,
\cite{Sacksteder2014} and \textit{Zitterbewegung},
\cite{Schliemann2006, Biswas2015} all of which have been derived
based on the assumption that the HH spin splitting is proportional
to $k^3$. Our work can be used to determine spin densities and
spin-Hall currents in the same way as for electrons.
\cite{Culcer_SpinCurrent_PRL07, Culcer_SteadyState_PRB07} It is also
highly relevant to the burgeoning field of hole quantum dots, which
are actively researched at present with a view to applications in
quantum computing, in particular via electric dipole spin resonance
(EDSR). \cite{Bulaev2007} The areal number densities of existing
single-hole quantum dots are contained in the parameter ranges we
study in this work in 2DHGs.

Apart from fully numerical calculations using an unconstrained basis
set, \cite{Broido1985, Winkler1992, Winkler1993} two approaches can
be adopted in regimes in which the cubic spin splitting
approximation is inadequate: One can evaluate higher-order terms in
perturbation theory, which quickly becomes cumbersome and
intractable, or one can perform a numerical diagonalization of the
effective Luttinger Hamiltonian restricted to a certain subspace
spanned by, e.g., the first and second HH and LH subbands. In this
work, we rely on the latter approach when the perturbative methods
fail. Our methods yield good agreement with fully numerical results
for GaAs holes. \cite{Broido1985, Winkler1992, Winkler1993}

Recently the importance of surface termination effects on spin-orbit
interactions and the HH spin splitting has been pointed out by
Durnev \textit{et al}, \cite{Durnev2014} who focused on the case of
quantum wells. Inclusion of these effects is beyond the scope of our
present work. Whereas in heterojunctions, where the wave functions
vanish near the interface, surface termination effects are expected
to be weaker,\cite{Durnev_private_comm_2016} a complete description
will require the systematic inclusion of these contributions.

This paper is structured as follows: In
Sec.~\ref{sec:TheorFormalism}, we introduce the variational method
employed for 2DHGs in quasitriangular wells. We apply this method
in Sec.~\ref{sec:RashbaBigSection} to calculate the spin splitting
and effective masses for 2DHGs in inversion and accumulation layers
in various semiconductors. The effects of the Dresselhaus spin-orbit
interaction terms are outlined in Sec.~\ref{sec:Dresselhaus}. We
discuss our results in Sec.~\ref{sec:Discussion}, while
Sec.~\ref{sec:Summary} contains a summary and conclusions.

\section{Theoretical Formalism}
\label{sec:TheorFormalism}

\subsection{Luttinger Hamiltonian for 2DHGs}
\label{subsec:LuttingerHamiltonian}

Our formulation takes as its starting point the bulk $4 \times 4$
Luttinger Hamiltonian \cite{Luttinger1956} describing holes in the
uppermost valence band with an effective spin $J=3/2$
\begin{equation} \label{eq:HLutt}
  H_L = \left(
  \begin{matrix}
    P + Q & 0 & L & M \\ 0 & P + Q & M^* & -L^* \\ L^* & M & P - Q &
    0 \\ M^* & -L & 0 & P - Q
  \end{matrix}
  \right),
\end{equation}
where
\begin{subequations}
  \label{eq:LuttingerMatrixElmts}
  \begin{align}
    P &= \frac{\mu}{2}\gamma_1 (k^2 + k_z^2), &
    Q &= -\frac{\mu}{2}\gamma_2
    \left( 2 k_z^2 - k^2 \right), \label{subeq: LuttElmts}\\
    L &= -\sqrt{3} \mu \, \gamma_3 k_- k_z, & M &= -\frac{\sqrt{3} \,
      \mu}{2} \left(\bar{\gamma} {k_-^2} - \zeta {k_+^2}
    \right),\\ \bar{\gamma} &= \frack{1}{2}\left(\gamma_2 + \gamma_3
    \right), & \zeta &= \frack{1}{2}\left(\gamma_3 - \gamma_2
    \right), \\ k^2 &= k_x^2 + k_y^2, & k_{\pm} &= k_x \pm i k_y,
  \end{align}
\end{subequations}
$\mu \equiv \hbar^2/m_0$ with bare electron mass $m_0$, and
$\gamma_1$, $\gamma_2$, and $\gamma_3$ are the Luttinger parameters,
see Table~\ref{tab:LuttingerParameters}. The wave vector components
are defined by the crystallographic orientation. In this work, we
consider holes grown on a (001) surface so that $k_x \parallel
(100)$, $k_y \parallel (010)$, and $k_z \parallel (001)$.  We have
expressed $H_L$ in the basis of $J_z$ eigenstates
$\{\ket{+\frac{3}{2}}, \ket{-\frac{3}{2}}, \ket{+\frac{1}{2}},
\ket{-\frac{1}{2}} \}$, where $\hat{\bm z}$ is the unit vector
perpendicular to the plane of the interface.

The Luttinger Hamiltonian is further simplified in the axial
approximation, where the terms proportional to $\zeta$ are
neglected. \cite{Winkler2003, Broido1985} The axial approximation is
appropriate for GaAs, InAs, InSb and Ge, while for Si $\zeta$ is
significant (Table~\ref{tab:LuttingerParameters}) and gives rise to
a highly anisotropic Fermi contour, as Sec.~\ref{subsec:Si} will
show.

\begin{table}
  \caption{\label{tab:LuttingerParameters} Luttinger parameters and
  bulk Dresselhaus coefficients \cite{Winkler2003} used in this
  work, where $B_{D1}$, $B_{D2}$, $B_{D3}$, and $B_{D4}$ are in
  eV{\AA}$^3$ and $C_D$ in eV{\AA}.}
  \begin{ruledtabular}
    \begin{tabular}{@{}>{$}c<{$}*{5}{d{2,4}}@{}}
      & \mc{GaAs} & \mc{InAs} & \mc{InSb} & \mc{Si} & \mc{Ge} \\ \hline
      \gamma_1 & 6.85 & 20.40 & 37.10 & 4.28 & 13.38 \\
      \gamma_2 & 2.10 &  8.30 & 16.50 & 0.34 &  4.24 \\
      \gamma_3 & 2.90 &  9.10 & 17.70 & 1.45 &  5.69 \\
      C_D      & -0.0034 & -0.0112 & -0.0082 & & \\
      B_{D1}   & -81.93 & -50.18 & -934.8 & & \\
      B_{D2}   & 1.47  & 1.26  & 41.73 & & \\
      B_{D3}   & 0.49  & 0.42  & 13.91 & & \\
      B_{D4}   & -0.98 & -0.84 & -27.82 & & \\
    \end{tabular}
  \end{ruledtabular}
\end{table}

The $4 \times 4$ Luttinger Hamiltonian (\ref{eq:HLutt}) is accurate
as long as the spin-orbit split-off band is far away from the HH and
LH bands. The energy gap $\Delta_\mathrm{SO}$ separating the
split-off band from the HH-LH manifold is of the order of
$300-800$~meV for GaAs, InAs, InSb, and Ge. \cite{Winkler2003} For
Si, $\Delta_\mathrm{SO} = 44$~meV, and thus the couplings to the
split-off band must be taken into account.

\subsection{Poisson and Schr\"{o}dinger Equations}
\label{subsec:Schroedinger-Poisson_Eqn}

We consider a single heterojunction with the interface at $z=0$.  We
assume that the wave functions vanish at the interface so that in
the following we restrict ourselves to $z\ge0$ (see end of Sec.~\ref{subsec:LinearAndCubicTerms} on why this is a reasonable approximation for the heterojunctions studied here). Our variational
calculation is based on two steps, each of which is variational in
nature. First we construct the self-consistent potential $V(z)$
characterizing the heterojunction assuming a parabolic, spin
degenerate, 2DHG dispersion. Then we solve $H_L$ for $V(z)$ in a
second variational calculation.

The one-dimensional charge distribution giving rise to the
confinement potential $V(z)$, consists of two contributions: the
hole density is $ p|\psi_{h}(z)|^{2} $, where $p$ denotes the number
density of 2D holes, and $\psi_{h}(z)$ is the zero-node HH wave
function, assuming that only the lowest subband, labeled HH1, is
occupied, which is the most common case.  The second contribution is
the net donor concentration $N_D$ (Ref.~\onlinecite{Stern1974}). The
corresponding Poisson equation is thus:
\begin{equation}
  \frac{d^2}{dz^2}V(z)= - \frac{e^2}{\epsilon_s\epsilon_0} \left[
    p|\psi_{h}(z)|^{2} + N_D \right],
  \label{eq:Poisson_eq}
\end{equation}
where $\epsilon_{s}$ is the dielectric constant of the semiconductor
and $\epsilon_{0}$ is the vacuum permittivity,

The Poisson equation (\ref{eq:Poisson_eq}) is solved with three
boundary conditions. Firstly, the zero of energy is chosen to be at
$z = 0$:
\begin{equation}
  V(z=0) = 0.
\end{equation}
Secondly, the potential $V(z)$ is flat in the bulk, i.e. at $z \geq
w$ we have
\begin{equation}
  \frac{d}{dz}V(z \geq w)= 0,
\end{equation}
where $w$ is the width of the space charge layer. \cite{Ando1982a}
Thirdly, the Fermi energy $E_F$ is given by the HH1 energy at the
Fermi wave vector $k_F$:
\begin{subequations}
\begin{equation}
  E_F \equiv E_{H1} + E_{F1},
\end{equation}
where $E_{H1}$ denotes the subband edge $k=0$ and $E_{F1}$ is the
in-plane kinetic energy of the holes in the HH1 subband at $E_F$
[i.e., $E_{F1} = \hbar^2 k_F^2 / (2m^\ast)$ in a system with a
simple parabolic dispersion characterized by an effective mass
$m^\ast$].  Furthermore, as the system is in equilibrium, $E_F$ is
constant throughout the system. Thus we have:
\begin{equation}
  V(z \geq w) = E_F + \Phi = E_{H1} + E_{F1} + \Phi
\end{equation}
\end{subequations}
where the band bending $\Phi$ is the energy difference between the
valence band edge and the Fermi energy in the bulk past the space
charge layer.

The solution of the Poisson equation (\ref{eq:Poisson_eq}), in the
Hartree approximation, \cite{Gross1991, Ando1982a} is given by:
\begin{equation}
  \label{eq:Conf_potential}
  V(z) = V_\mathrm{2DHG} (z) + V_D(z)
\end{equation}
where $V_\mathrm{2DHG}(z)$ is the electrostatic potential due to the
2D holes occupying the HH1 subband and $V_D(z)$ is due to the
charged donors in the space charge layer. \cite{Broido1985, 
Ando1982a} Assuming that the wave functions $\psi_h(z)$ vanish at
the interface $z=0$, the term $V_\mathrm{2DHG}(z)$ becomes
\begin{equation}
  \label{eq:V2DHG}
  V_\mathrm{2DHG} (z) =\frac{pe^2}{\epsilon_s \epsilon_0} \left[
    z - \int_{0}^{z} dz' \int_{0}^{z'} dz'' |\psi_h(z'')|^2
  \right] 
\end{equation}
whereas the contribution from the space charge layer becomes
\begin{equation}
  V_D(z) =
  \frac{e^2}{\epsilon_{s}\epsilon_0}N_D\left(w z - z^{2}/2 \right),
  \label{eq:V_imp}
\end{equation}
where
\begin{equation}
  w \equiv \sqrt{\frac{2 \epsilon_s \epsilon_0}{e^2
      N_D} \left[\Phi + E_{H1} + E_{F1} -
      \frac{e^2}{\epsilon_s \epsilon_0} p \braket{z} \right]},
  \label{eq:w_depl}
\end{equation}
and the expectation value $\braket{z}$ is defined via the wave function
$\psi_h (z)$.

The wave function $\psi_{h}(z)$ entering the Poisson equation
(\ref{eq:Poisson_eq}) is the solution of the Schr\"odinger equation.
Thus one usually solves the Poisson and Schr\"{o}dinger equations in
a self-consistent iterative scheme. \cite{Broido1985, Bangert1985,
  Ekenberg1984, Volkov1998, Winkler2002, Ando1982} Using $V(z)$ from
the Poisson equation, the 2D hole density $p|\psi_{h}(z)|^{2}$ is
obtained from the Hamiltonian, from which one then constructs a new
$V(z)$ by solving the Poisson equation. The process is iterated
until $V(z)$ converges. In this work, we instead employ a simplified
procedure based on the self-consistent variational scheme presented
in Ref.~\onlinecite{Winkler1992}, which yields good agreement with
fully self-consistent numerical calculations.  We approximate the
wave functions using Fang-Howard variational wave functions, which
take the form \cite{Fang1966}
\begin{equation}
  \psi_{v}(z)=2\lambda_v^{3/2}z\exp(-\lambda_v z)
  \label{eq:FHGroundState-1}
\end{equation}
where $v=h$ represents the zero-node HH1 wave function.  For the
wave functions entering $V(z)$ we neglect $k$-dependent band mixing,
which has only a small effect on $V(z)$.  The variational parameter
$\lambda_{h}$ is obtained by minimizing the $k=0$ ground state HH
energy $E_{H1}$, which, neglecting band mixing, is the sum of the
diagonal matrix element of the Luttinger Hamiltonian in Eq.\
(\ref{eq:HLutt}) for the HH subspace and the expectation value of
$V(z)$ in Eq.\ (\ref{eq:Conf_potential}), taking into account that
$E_{H1}$ also appears in Eq.\ (\ref{eq:w_depl}).

In the second variational step, we obtain the $\bm{k}$ and spin
dependent eigenfunctions $\Psi_{h\bm{k}} (z)$ of the total
Hamiltonian $\tilde{H} = H_L + V(z)$ for the HH1 sub-band by
expanding $\Psi_{h\bm{k}} (z)$ in terms of the lowest eigenstates of
$\tilde{H}$ for $\bm{k}=0$, when $\tilde{H}$ becomes diagonal. For
the $\bm{k}=0$ HH1 and LH1 states we use the zero-node wave
functions (\ref{eq:FHGroundState-1}) with $v = h,l$.  For the
one-node HH2, LH2 states we use the form \cite{Mori1979, Kelly1990}
\begin{multline}
  \label{eq:FHExState}
  \psi _w (z) = \sqrt{12} \lambda_w^{3/2} z \left[1-\left(\lambda_v
    + \lambda_w \right)z/3\right]\\ \times e^{-\lambda_w z} /
  \sqrt{1 - \lambda_v/\lambda_w + \lambda_v^2/\lambda_w^2}.
\end{multline}
By construction, these wave functions are orthogonal to the
zero-node wave functions (\ref{eq:FHGroundState-1}).  The quantities
$\lambda_w$ with $w = H,L$ denote additional variational
parameters. The introduction of these additional variational
parameters improves the accuracy of the method \cite{Kelly1990} as
compared to using a single variational parameter $\lambda_v =
\lambda_w$.
The eigenvalues $E(\bm{k})$ and the corresponding $\bm{k}$ dependent
expansion coefficients are obtained by diagonalizing the matrix
$\tilde{H}$, whose elements are given as
\begin{equation}
  \tilde{H}_{\nu\nu'}=\langle \nu | H_L + V(z) \ket{\nu'},
  \label{eq:subband_HLutt}
\end{equation}
where $|\nu\rangle$ denotes the wave functions
(\ref{eq:FHGroundState-1}) and (\ref{eq:FHExState}). The two lowest
eigenenergies of the $8\times 8$ matrix (\ref{eq:subband_HLutt})
correspond to the dispersion of the spin-split HH1$_{\pm}$
subband. In certain regimes these eigenenergies can also be obtained
analytically to a good approximation, as shown in the next section.

\subsection{Schrieffer-Wolff Transformation and Rashba Spin Splitting}
\label{subsec:SW}

\setlength{\tabcolsep}{6pt}
\begin{table*}
  \caption{\label{tab:Coefficients} Values for the material- and
    structure-dependent coefficients $A$ (in
    $10^{-16}$~meV$\,$m$^{2}$), $B$ (in $10^{-32}$~meV$\,$m$^{4}$),
    and $C\equiv|\alpha_{R2}|$ (in $10^{-24}$~meV$\,$m$^{3}$) in the
    dispersion relation $E(k) = A k^2 - Bk^4 \pm C k^3 $, the
    energies $E_{H1}$, $\Delta_{11}^{HL}$, $E_{F1}\equiv
    E_{F}-E_{H1}$ (in meV), and spin splitting $\Delta p$ for GaAs
    inversion and accumulation layers with $N_{D}-N_{A} =
    3\times10^{20}$~m$^{-3}$. The densities $p$ and $\Delta p$ are in multiples of
    $10^{15}$~m$^{-2}$. For inversion layers, the Schrieffer-Wolff approximation fails
    when the density exceeds $2.5\times10^{15}$~m$^{-2}$. For
    accumulation layers it is valid up to a density of $0.5\times10^{15}$~m$^{-2}$.}

  \begin{ruledtabular}
    \begin{tabular}{ccccccccc}
      Density $p$ & $A$ & $B$ & $C$ & $E_{H1}$ & $\Delta_{11}^{HL}$ &
      $E_{F1}$ &$\Delta p$%
      \footnote{Calculated using the variational method introduced in Sec.~\ref{subsec:Schroedinger-Poisson_Eqn} and a numerical
        diagonalization of Eq.\ (\ref{eq:subband_HLutt}).}  &$\Delta
      p$%
      \footnote{Calculated using the fully numerical method devised in
        Refs.~\onlinecite{Winkler1992, Winkler1993}.}  \\[0.08cm]
		\hline & \multicolumn{8}{c}{Inversion layer} \\ 
		0.5 &2.99 &0.27 &0.53 &16.12 &8.32 &0.91 &0.10 &0.14 \\
		1.0 &2.92 &0.25 &0.58 &19.24 &8.73 &1.67 &0.18 &0.23 \\ 
		1.5 &2.86 &0.23 &0.61  &22.12 &9.04 &2.30 &0.25 &0.30 \\ 
		2.0 &2.80 &0.22 &0.64 &24.83  &9.27 &2.83 &0.31 &0.35 \\ 
		2.5 &2.74 &0.20 &0.67  & 27.40 &9.46 &3.26 &0.36 &0.39 \\ 
		3.0 & & & & 29.85 &9.62 &3.64 &0.41 &0.42
		\\ \hline & \multicolumn{8}{c}{Accumulation layer} \\ 
		0.5  &2.55 &0.69 &1.15 &8.84 &2.68 &0.70 &0.31 &0.33 \\ 
		1.0 & & &  &12.85 &2.87 &1.06 &0.48 &0.43 \\ 
		1.5 & & & &16.30 &2.97 &1.36 &0.55 &0.47 \\ 
		2.0 & & & &19.42 &3.04 &1.66 &0.58  &0.49 \\ 
		2.5 & & & &22.31 &3.08 &1.96 &0.60 &0.50 \\ 
		3.0 & & & &25.02 &3.12 &2.28 &0.60 &0.51
    \end{tabular}
  \end{ruledtabular}
\end{table*}

It is well established that, under certain circumstances, the HH1
subbands may be described by an effective $2\times 2$ Hamiltonian
formulated as an expansion in powers of the wave vector
$\bm{k}$. \cite{Winkler2003} The general kinematic structure of such
a reduced Hamiltonian can be found from the theory of
invariants. \cite{Bir74} Retaining terms up to fourth order in
$\bm{k}$ and postponing the discussion of Dresselhaus terms until
Sec.~\ref{sec:Dresselhaus}, the effective $2\times 2$ Hamiltonian
for the subspace spanned by the spin-split HH1 subbands takes the
form \cite{Winkler2003}
\begin{equation}
  \begin{array}[b]{rcL}
  H_{2\times2} & =
  & [A k^2 - B k^4 - d(k_+^2 - k_-^2)^2] \openone_{2\times2} \\[0.5ex]
  & & + i\alpha_{R1}(k_+\sigma_+ - k_-\sigma_-)
  + i\alpha_{R2}(k_+^3\sigma_- - k_-^3\sigma_+) \\[0.5ex]
  & & + i\alpha_{R3}(k_+\sigma_+ - k_-\sigma_-) k^2 \\[-2ex]
  \end{array}
  \label{eq:2x2_Hamiltonian_with_Rashba}
\end{equation}
Up to fourth order in $\bm{k}$, the Hamiltonian $H_{2\times2}$
includes all possible terms.  The terms weighted by $A$, $B$, and
$d$ describe the orbital motion, whereas the terms weighted by
$\alpha_{Ri}$ represent the Rashba spin-orbit coupling.  More
specifically, the term $Ak^2$ describes the usual parabolic
component of the dispersion, while the term $Bk^4$ represents the
lowest-order nonparabolic correction, which remains isotropic and
often has a sizable effect on the dispersion of hole systems, as
discussed below. The term $d(k_+^2 - k_-^2)^2]$ characterizes the
warping of the energy contours. As can be seen from
Eq.\ (\ref{eq:2x2_Hamiltonian_with_Rashba}), the most general form
of the Rashba SO coupling includes a term linear in $k$
proportional to $\alpha_{R1}$ and two terms cubic in $k$ weighted
by $\alpha_{R2}$ and $\alpha_{R3}$. The terms weighted by $d$,
$\alpha_{R1}$, and $\alpha_{R3}$ break axial symmetry, which
implies that these prefactors are zero when the axial
approximation $\zeta = 0$ is employed in
Eq.~(\ref{eq:HLutt}). Moreover, to lowest order the prefactor
$\alpha_{R1}$ stems from the $\bm{k}\cdot\bm{p}$ coupling between
the bonding and anti\-bonding atomic $p$ orbitals (the latter give
rise to the first excited conduction band). \cite{Winkler2003} For
the systems discussed here, the $k$-linear Rashba term thus
contributes not more than $\sim 1$\% of the total spin splitting,
consequently this term is not considered further.

To obtain analytical expressions for the prefactors appearing in
Eq.\ (\ref{eq:2x2_Hamiltonian_with_Rashba}), one can apply a
Schrieffer-Wolff transformation \cite{Schrieffer1966} (L\"{o}wdin
perturbation theory \cite{Lowdin1951}) to
Eq.\ (\ref{eq:subband_HLutt}).  The coefficients $A$, $B$, $d$,
$\alpha_{R2}$ and $\alpha_{R3}$ are evaluated to lowest order in the
perturbation expansion: Explicit expressions are given in the
Appendix. In the axial approximation, with $d = \alpha_{R1} =
\alpha_{R3} = 0$, the dispersion relation for the HH1 subband takes
the simple form
\begin{equation}
  E_\pm (k) = A k^2 - Bk^4 \pm C k^3,
  \label{eq:SW_dispersion_with_Rashba}
\end{equation}
where $C\equiv|\alpha_{R2}|$. Numerical values for $A, B,$ and
$C\equiv|\alpha_{R2}|$ for typical experimental densities are given
in Table~\ref{tab:Coefficients}. The resulting subbands, denoted as
HH1+ and HH1$-$, have unequal subband populations $p_{\pm}$ and
density of states (DOS) effective masses $m_{\pm}$.

To characterize the strength of the Rashba spin-orbit
interaction, we use the dimensionless quantity
\begin{equation}
  \label{eq:delta_p}
  \Delta p \equiv \frac{|p_+-p_-|}{p}
\end{equation}
where $p_\pm$ denotes the spin subband densities with
$p = p_+ + p_-$. Experimentally, the quantity $\Delta p$ is usually
inferred by analyzing the beating pattern of Shubnikov-de-Haas (SdH)
oscillations. \cite{Ando1982, Winkler2003, Nichele2014a, Grbic2007, 
Liles2016} For the single heterojunctions studied here, it can be
manipulated by tuning the density $p$. At low temperature, $p_\pm$
is given by
\begin{equation}
  p_{\pm} = \int \frac{ d^2 k}{(2\pi)^2} \theta [E_F - E_{\pm}(k)] .
  \label{eq:p_plus_minus_general}
\end{equation}
where $\theta$ is the Heaviside step function. For hole systems with
isotropic Fermi contours, Eq.\ (\ref{eq:p_plus_minus_general})
becomes
\begin{equation}
  p_{\pm} = \frac{k_{F\pm}^2}{4\pi}.
  \label{eq:p_plus_minus_isotropic}
\end{equation}
where $k_{F\pm}$ denotes the Fermi wave vectors for the spin-split bands
Using the coefficients in Eq.\ (\ref{eq:SW_dispersion_with_Rashba}),
this translates to
\begin{equation}
  \label{eq:p_plus_minus_p}
  p_{\pm} =\frac{p}{2}\pm
  \frac{pc}{\sqrt{2}X}\sqrt{p\pi\left(6-\frac{4}{X}\left(1-4p\pi
    b\right)\right)}
\end{equation}
where
\begin{subequations}
  \begin{align}
    X &\equiv 1 - 4p\pi b + \sqrt{1 - 4p\pi(2b + c^2 -
      4b^2p\pi)}\\[1ex] b &\equiv B/A \hspace{2em} c \equiv C/A
  \end{align}
\end{subequations}
so that the Rashba spin splitting $\Delta p$ is given by
\begin{equation}
  \Delta p =
  \frac{\sqrt{2}c}{X}\sqrt{p\pi\left(6-\frac{4}{X}\left(1-4p\pi
    b\right)\right)}.
\end{equation}

The DOS effective masses $m_{\pm}$ of the spin-split subbands at the
Fermi energy $E_F$ takes the form
\begin{equation}
  \frac{m_{\pm}}{m_0} = \frac{\mu}{2\pi} \int d^2k \, \delta[E_F -
    E_{\pm}(k)],
\end{equation}
where $\delta$ is the Dirac $\delta$ function. For isotropic bands,
this becomes
\begin{subequations}
  \label{eq:Eff_mass_isotropic}
  \begin{equation}
    \frac{m_{\pm}}{m_0} = \mu
    \left(\frac{1}{k}\frac{dE_\pm(k)}{dk}\right)^{-1}_{k=k_{F\pm}},
  \end{equation}
  which can be further evaluated for the dispersion
  (\ref{eq:SW_dispersion_with_Rashba})
  \begin{equation}
    \frac{m_{\pm}}{m_0} = \frac{\mu}{2A - 4Bk_{F\pm}^2 \pm 3Ck_{F\pm}},
  \end{equation}
\end{subequations}
where the Fermi wave vectors $k_{F\pm}^2$ for a given total density
$p$ are evaluated using Eqs.\ (\ref{eq:p_plus_minus_isotropic}) and
(\ref{eq:p_plus_minus_p}).

\subsection{Comparison with Numerical Results}
\label{subsec:ComparisonNum}

\begin{figure}
  \includegraphics[scale=0.65]{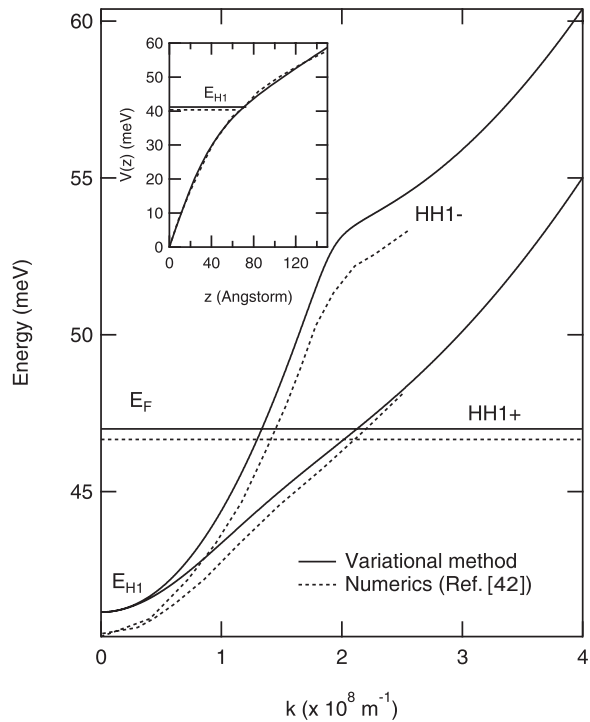}
  \caption{Comparison between the variational method adopted
    in this work (solid lines) and the numerical results in
    Ref.~\onlinecite{Broido1985} (dotted lines), showing the
    inversion-asymmetric potential (inset) and the spin-split HH1+
    and HH1$-$ subbands. The dispersion was obtained by a numerical
    diagonalization of Eq.\ (\ref{eq:subband_HLutt}). The system is a
    GaAs single heterojunction (SHJ) with a 2D hole density $p =
    5\times10^{15}$~m$^{-2}$ and a net dopant concentration of
    $N_D-N_A = 1\times10^{21}$ m$^{-3}$.}
  \label{fig:dispersion_cf_Broido}
\end{figure}

To illustrate our approach, Fig.~\ref{fig:dispersion_cf_Broido}
shows a comparison between the results obtained using our
variational calculation and those obtained in
Ref.~\onlinecite{Broido1985} using an iterative Fang-Howard and
Luttinger Hamiltonian scheme. The 2DHG considered in
Ref.~\onlinecite{Broido1985} is a hole GaAs single heterojunction
with a density of $5\times10^{15}$~m$^{-2}$ and a net dopant
concentration of $N_D-N_A = 1\times10^{21}$~m$^{-3}$.
\cite{Stormer1983} The confinement potential $V(z)$ obtained from
solving Eq.\ (\ref{eq:Poisson_eq}) is shown in the inset of
Fig.~\ref{fig:dispersion_cf_Broido}. Our variational approach
slightly overestimates the ground state heavy hole energy $E_{H1}$
compared to the numerical results, as expected for a variational
calculation. The corresponding DOS effective masses, obtained from
Eqs.\ (\ref{eq:subband_HLutt}) and (\ref{eq:Eff_mass_isotropic}),
are $m_+=0.52~m_0$ and $m_-=0.12~m_0$. These numbers are in close
agreement with the full-numerical calculations in
Ref.~\onlinecite{Broido1985}, where $m_+=0.46~m_0$ and $m_- =
0.12~m_0$. We also find good agreement between our variational
results for the Rashba spin splitting $\Delta p$ and those obtained
using the numerical method described in
Refs.~\onlinecite{Winkler1992, Winkler1993} (see
Table~\ref{tab:Coefficients}).

While the simple dispersion in
Eq.\ (\ref{eq:SW_dispersion_with_Rashba}) affords a convenient way
to calculate DOS properties such as effective masses and spin
splittings, the validity of the Schrieffer-Wolff transformation is
limited to certain sets of densities and dopant concentrations. When
the separation between subband energies is small, the coefficients
$A, B$ and $C$ are overestimated (refer to the Appendix for the
dependence of $A, B$ and $C$ on the energy separations) and one can
no longer describe the HH1+ and HH1$-$ bands by
Eq.\ (\ref{eq:SW_dispersion_with_Rashba}). For example, using the
values in Table~\ref{tab:Coefficients}, for a GaAs inversion layer
with $p = 2.5\times10^{15}$~m$^{-2}$ and $N_{D}-N_{A} =
3\times10^{20}$~m$^{-3}$, the Rashba term $C k^3 $ becomes larger
than $A k^2 - B k^4$ so that the heavier HH1+ band bends down for $k
> k_\mathrm{bend} = 1.7\times 10^{8}$~m$^{-1}$. As $k_{F+}=1.5
\times10^{8}$~m$^{-1}$ is very close to HH1+ turning point
$k_\mathrm{bend}$, the HH1+ dispersion is almost flat at $k_{F+}$,
which means that the HH1+ effective mass is overestimated. The
Schrieffer-Wolff results for GaAs inversion layers with $N_{D}-N_{A}
= 3\times 10^{20}$~m$^{-3}$ and $p \gtrsim
2.5\times10^{15}$~m$^{-2}$ are invalid and we resort to a numerical
diagonalization of Eq.\ (\ref{eq:subband_HLutt}) to obtain the
energy dispersion. Note that unlike
Eq.\ (\ref{eq:SW_dispersion_with_Rashba}), dispersion curves
obtained from a numerical diagonalization of
Eq. (\ref{eq:subband_HLutt}) include spin splitting to all orders in
$k$. Therefore, the dispersions from Eq.\ (\ref{eq:subband_HLutt})
are a reasonable approximation for any wave number $k$ (and hence
for any density $p$), so long as only the zero-node heavy hole
subbands HH$1\pm$ are occupied. Indeed, in all the structures
discussed in this paper, at the densities we consider, only the HH1+
and HH1$-$ bands are occupied (Tables \ref{tab:Coefficients},
\ref{tab:EnergyLevelSpacingsDiffMaterials}, and
\ref{tab:EnergyLevelSpacingsInSb}).

\section{Full Rashba Spin-Splittings and Effective Masses}
\label{sec:RashbaBigSection}

Generally, spin splittings for holes are much larger than for
electrons: For electron inversion layers with the same doping
concentration, the spin splittings are about two orders of magnitude
smaller. \cite{Winkler2003} Below we focus first on GaAs inversion
and accumulation layers, which are distinguished by the position of
the Fermi level $E_F$ towards the subtrate: in an inversion
(accumulation) layer, $E_F$ is pinned near the conduction (valence)
band. Finally, we discuss the spin splitting and effective masses in
InAs, InSb, Ge, and Si hole inversion layers.

\subsection{GaAs Inversion and Accumulation Layers}
\label{subsec:Inv_vs_acc_GaAs}

\begin{figure}
  \includegraphics[scale=0.65]{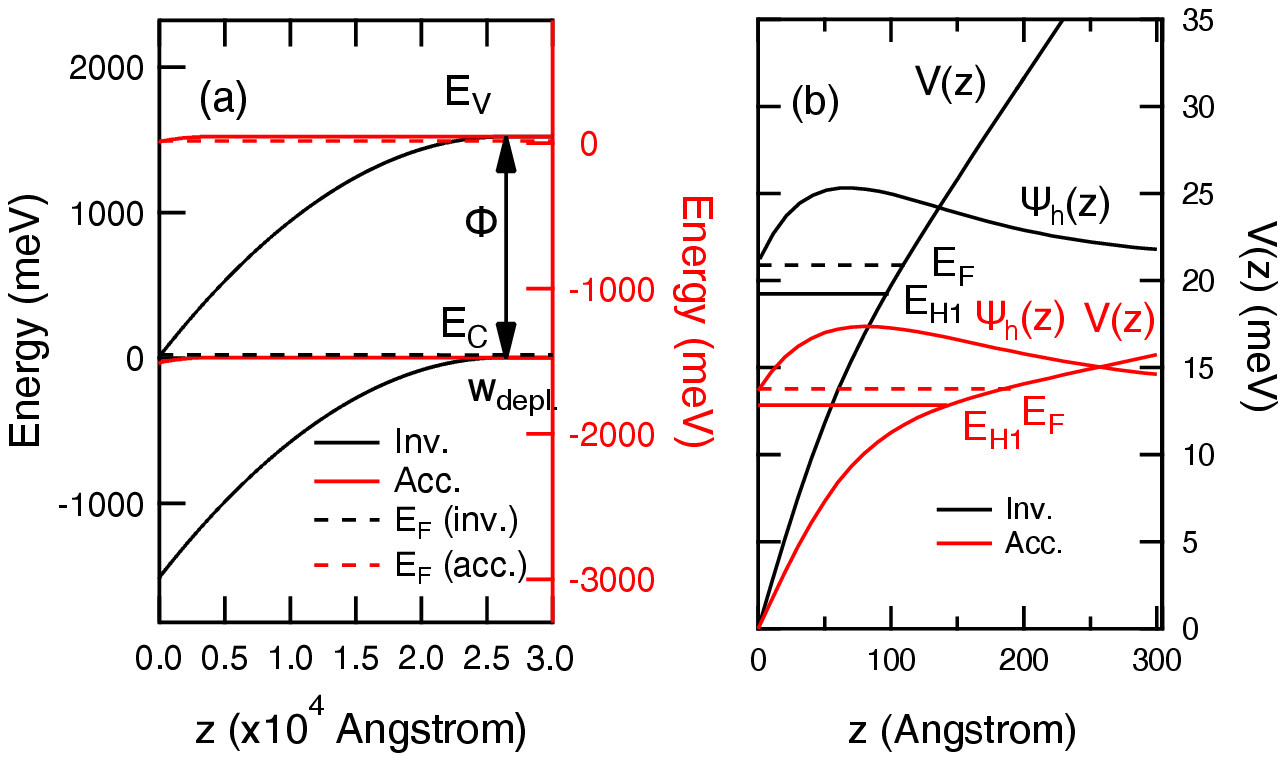}
  \caption{(Color online) Comparison of the (a) band structure and
    (b) confinement potential for a GaAs 2D hole system in an
    inversion and an accumulation layer with a density of
    $1\times10^{15}$~m$^{-2}$, where the net dopant concentration is
    $N_D-N_A = 3\times10^{20}$ m$^{-3}$ and $N_D =
    3\times10^{20}$~m$^{-3}$, respectively. Due to the magnitude of the
    band bending, the potential for an inversion layer is steeper
    than that of an accumulation layer.}
  \label{fig:pot_inv_vs_acc}
\end{figure}

The type of background dopant determines the location of the Fermi
level $E_F$ in the substrate and hence the amount of valence band
bending $\Phi$ at the heterojunction interface
[Fig.~\ref{fig:pot_inv_vs_acc}(a)]. For an accumulation layer, the
band bending is less pronounced than for an inversion
layer. Consequently, in accordance to Gauss' law, for the same
density, the confinement potential for an accumulation layer is less
steep than in an inversion layer
[Fig.~\ref{fig:pot_inv_vs_acc}(b)]. This means that the spacing
between subbands is smaller (Fig.~\ref{fig:inv_vs_acc_dispersion}
and Table~\ref{tab:Coefficients}), hence, according to the
expressions in the Appendix, the Rashba SO interaction is stronger
in an accumulation layer.

\begin{figure}
  \includegraphics[scale=0.7]{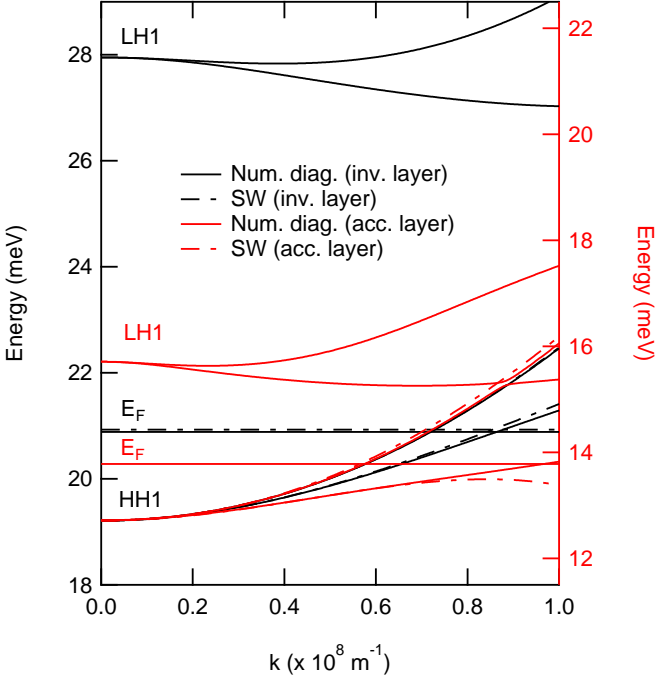}
  \caption{(Color online) Dispersion for 2D holes in a GaAs
    inversion layer (black lines) and a GaAs accumulation layer (red
    lines) with a density of $1\times10^{15}$~m$^{-2}$. The doping
    concentration is $N_D-N_A = 3\times10^{20}$~m$^{-3}$ for the
    inversion layer and $N_D = 3\times10^{20}$~m$^{-3}$ for the
    accumulation layer. The SW dispersion closely matches the
    numerical diagonalization results (black solid lines) up to
    $k_F$ for the inversion layer (black dash-dotted lines). In the
    accumulation layer, the HH1-LH1 separation is closer than in the
    inversion layer, so that the HH1-LH1 anticrossing occurs at a
    lower $k$ compared to the inversion layer, and the Schrieffer-Wolff method
    fails at this density (red dash-dotted lines).}
  \label{fig:inv_vs_acc_dispersion}
\end{figure}

\begin{figure}
  \includegraphics[scale=0.75]{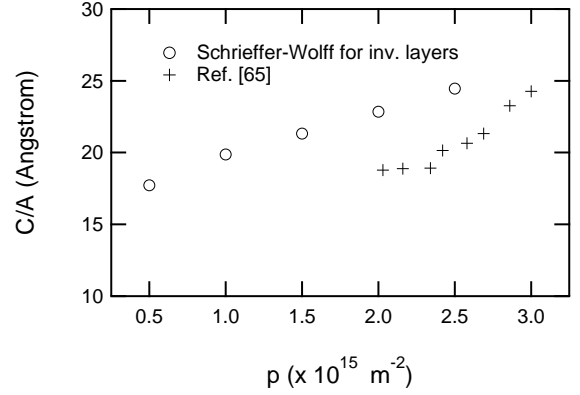}
  \caption{The ratio of $C$ to $A$ in the dispersion relation of
    $E(k) = A k^2 - Bk^4 \pm C k^3 $ for GaAs inversion layers with
    a dopant concentration of $3\times10^{20}$~m$^{-3}$. The trend
    we predict agrees with the experimental results in
    Ref.~\onlinecite{Grbic2007}.}
  \label{fig:ratio_of_C_to_A_GaAs}
\end{figure}

The variational approach followed in this work is designed to yield
the energies of the HH1$\pm$ subbands. Although it is not intended
to give reliable values for the higher subbands, some qualitative
observations can be made concerning the HH1-LH1 spacing in inversion
and accumulation layers. In Table~\ref{tab:Coefficients} and
Fig.~\ref{fig:inv_vs_acc_dispersion}, the HH1-LH1 separation
$\Delta_{11}^{HL}\equiv |E_{H1}-E_{L1}|$ at $k=0$ is smaller in an
accumulation layer than in an inversion layer at the same density
(Table~\ref{tab:Coefficients}). Since the Rashba coefficient
increases as the subbands get closer together, it is larger in an
accumulation than an inversion layer at a given density (see
Appendix for the dependence of the Rashba coefficient on the subband
energy separations). The Rashba coefficient increases with density,
which is consistent with the experimental results reported in
Ref.~\onlinecite{Grbic2007} (Fig.~\ref{fig:ratio_of_C_to_A_GaAs}).

\begin{figure}
  \includegraphics[scale=0.75]{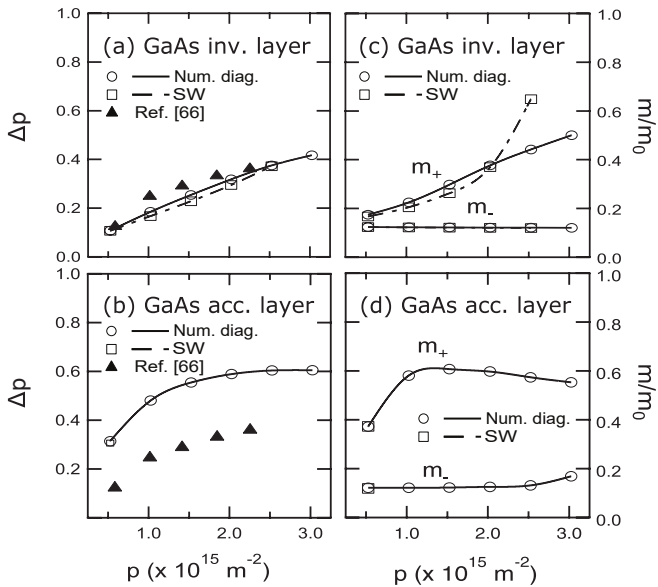}
  \caption{Rashba spin splitting $\Delta p$ for holes in GaAs
    (a) inversion layers and (b) accumulation layers compared to the
    experimental results reported in Ref.~\onlinecite{Liles2016}. The
    saturation of the spin splitting in accumulation layers is due
    to the HH1-LH1 anticrossing. In (c) we show the effective
    masses $m_{\pm}$ for GaAs inversion layers and in (d) for
    accumulation layers. The heavier mass $m_+$ increases with
    density whereas $m_-$ is nearly density-independent. In
    accumulation layers $m_+ $ saturates with density due to
    the proximity of $k_{F-}$ to the HH1-LH1 anticrossing. The
    doping concentration is $N_D-N_A = 3\times10^{20}$~m$^{-3}$ for
    the inversion layers and $N_D = 3\times10^{20}$~m$^{-3}$ for the
    accumulation layers.}
  \label{fig:inv_vs_acc_eff_masses}
\end{figure}

We compare the different trends in the Rashba spin splitting
$\Delta p$ as a function of density in inversion and accumulation
layers in Fig.~\ref{fig:inv_vs_acc_eff_masses}(a) and (b).  As
expected, $\Delta p$ increases with density in both inversion and
accumulation layers, consistent with the experimental observations
of Ref.~\onlinecite{Liles2016}. However, there is a difference
between the dependence of $\Delta p$ on density for the inversion
and accumulation layers.  For inversion layers with
$N_D-N_A = 3\times10^{20}$~m$^{-3}$ and $p$ ranging from
$5\times10^{14}$ m$^{-2}$ to $3\times10^{15}$~m$^{-2}$, the Rashba
spin splitting increases with density in an almost linear
fashion. In the accumulation layer counterparts, however, the spin
splitting increases in an almost linear fashion at lower densities
but saturates at higher densities. This feature can be attributed to
the fact that, in an accumulation layer, the HH1-LH1 separation is
smaller than in an inversion layer (see Appendix) such that the
HH1$-$ band anticrosses with the next highest energy subband (LH1)
at a lower $k$ than in an inversion layer at the same density
(Fig. {\ref{fig:inv_vs_acc_dispersion}}). Consequently, for an
accumulation layer with a higher density, $k_{F-}$ can be near the
HH1-LH1 anticrossing.  In the anticrossing region the HH1$-$ band
is pushed down in energy, hence the separation between the HH1$-$
and HH1+ bands is reduced.

\begin{figure}
 \includegraphics[scale=0.75]{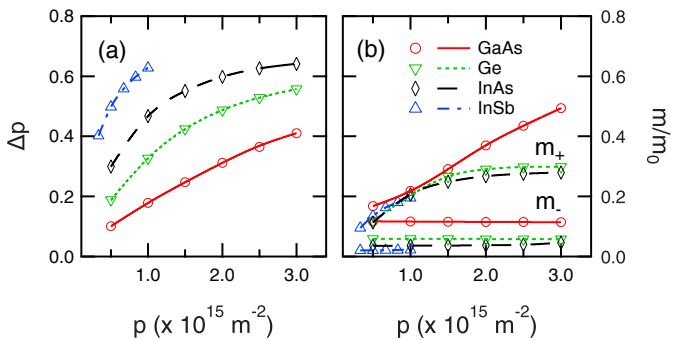}
  \caption{(Color online) (a) Rashba spin splitting $\Delta p$ and
    (b) effective masses $m_{\pm}$ for various inversion
    layers. The results are obtained from a numerical diagonalization of
    Eq.\ (\ref{eq:subband_HLutt}), with $N_D-N_A = 3\times10^{20}$~m$^{-3}$.}
  \label{fig:spin-splitting_and_eff_masses}
\end{figure}

The strength of the Rashba SOI is also evident in the difference
between the HH1+ and HH1$-$ effective
masses. Figures \ref{fig:inv_vs_acc_eff_masses}(c) and
\ref{fig:inv_vs_acc_eff_masses}(d) show the variation of the
effective HH1+ and HH- masses with density. As the figures show,
there is a remarkable distinction in the dependence of the $m_+$ and
$m_-$ on density for the inversion and accumulation layers we
consider. For the inversion layers studied here, the HH1+ effective
mass also increases in an almost linear fashion with density,
whereas $m_-$ is essentially constant. The increase in the
difference between the effective masses $m_+$ and $m_-$ with density
again implies that the strength of the Rashba SO interactions
increase with density. Comparison between numerical and SW results
for inversion layers with densities up to $p = 2 \times
10^{15}$~m$^{-2}$ shows the SW approach works well. However, at a
density $p=2.5\times10^{15}$ m$^{-2}$, the HH1+ effective mass,
calculated using the values in Table~\ref{tab:Coefficients}, is
overestimated [Fig.~\ref{fig:inv_vs_acc_eff_masses}(c)], which
signifies that SW breaks down. For the accumulation layers
considered here, however, with densities ranging from
$5\times10^{14}$~m$^{-2}$ to $3\times10^{15}$~m$^{-2}$, the HH1+
effective mass increases, then saturates, and becomes increasingly
lighter as the density increases. The SW approach is only valid for
densities up to $\sim5\times10^{14}$~m$^{-2}$ for the accumulation
layers. That SW fails at lower densities in accumulation layers is
expected as the separation between subband energies
in an accumulation layer is smaller than in an inversion layer at a
given density. On the other hand, the HH1$-$ effective mass is
essentially constant at lower densities but increases slightly at a
higher density of $p=3\times10^{15}$ m$^{-2}$
[Fig.~\ref{fig:inv_vs_acc_eff_masses}(d)]. The dependence of the
HH1+ effective mass $m_+$ on density can be explained by examining a
typical accumulation layer band structure
(Fig.~\ref{fig:inv_vs_acc_dispersion}). As the density increases,
the Fermi energy increases, and the fact that the HH1+ band
curvature becomes slightly steeper at $k > 1 \times 10^{8}$~m$^{-1}$
means that the HH1+ effective mass at the Fermi energy decreases
slightly with increasing density. The behavior of HH1$-$, on the
other hand, reflects the anticrossing between the HH1-LH1 bands. As
explained above, at a sufficiently high density, $k_{F-}$ can be
very close to the HH1-LH1 anticrossing. In this region, the HH1$-$
band becomes flatter, hence the corresponding effective mass $m_-$
increases slightly.

\subsection{Zincblende Materials and Ge Inversion Layers}
\label{subsec:Zincblende_and_Ge}

Figure~\ref{fig:spin-splitting_and_eff_masses} shows the spin
splitting $\Delta p$ and effective masses $m_{\pm}$ as functions of
density for inversion layers in GaAs, Ge, InSb, and InAs. The
strength of the SO interaction is reflected in the difference
between the subband populations and effective masses of the HH1+ and
HH1$-$ subbands. In Ge, InAs, and InSb, the Rashba spin splitting
and HH1+ effective mass $m_+$ saturate at a lower density compared
to GaAs. The saturation of the spin splitting and $m_+$ indicates
that HH1 and LH1 are close enough such that the HH1-LH1
anticrossing occurs near $E_{F}$. This indicates that Ge, InSb, and
InAs exhibit a stronger SO interaction than GaAs, which is
consistent with the known fact that SO coupling is stronger in
compounds containing elements with larger atomic numbers $Z$. The
corollary of this is that SW breaks down at different densities for
various materials, which implies that
Eq.\ (\ref{eq:SW_dispersion_with_Rashba}) is valid only for low
densities for heavy materials. For example, using the values for
$A$, $B$, and $C$ listed in
Table~\ref{tab:CoefficientsDiffMaterials}, one can get a reasonable
estimate for the HH1+ and HH1$-$ effective masses of InAs and Ge at
a density of $5\times10^{14}$~m$^{-2}$. The SW effective masses of
InAs are obtained as $m_+ = 0.094~m_0$ and $m_- = 0.036~m_0$, which
are close to $m_+ = 0.113~m_0$ and $m_- = 0.036~m_0$ obtained by
numerically diagonalizing Eq.\ (\ref{eq:subband_HLutt}). For Ge, the
SW effective masses are $m_+ = 0.108~m_0$ and $m_- = 0.059~m_0$,
which are in excellent agreement with the numerical diagonalization
results $m_+ = 0.118~m_0$ and $m_- = 0.059~m_0$. For InSb, the SW
results are valid at densities $\lesssim
4 \times 10^{14}$~m$^{-2}$. Using the values in Table
\ref{tab:EnergyLevelSpacingsInSb} for $3.3 \times 10^{14}$~m$^{-2}$,
the SW effective masses are $m_+ = 0.087~m_0$ and $m_- = 0.019~m_0$,
which also compare well with the numerical diagonalization results
of $m_+ = 0.095~m_0$ and $m_- = 0.020~m_0$.

\begin{table*}
  \caption{\label{tab:CoefficientsDiffMaterials} Values for
    material-dependent constants $A$ (in $10^{-16}$~meV$\,$m$^{2}$),
    $B$ (in $10^{-32}$~meV$\,$m$^{4}$), and $C$ (in
    $10^{-24}$~meV$\,$m$^{3}$) in the dispersion relation $E(k) = A
    k^2 - Bk^4 \pm C k^3 $ for GaAs, InAs, and Ge inversion layers
    with $N_{D}-N_{A} = 3 \times 10^{20}$~m$^{-3}$. All densities below are in
    $10^{15}$~m$^{-2}$. The Schrieffer-Wolff approach breaks down for densities exceeding
    $0.5 \times 10^{15}$~m$^{-2}$ for InAs as well as for Ge.}

  \begin{ruledtabular}
    \begin{tabular}{c|ccc|ccc|ccc}

      & \multicolumn{3}{c|}{GaAs} & \multicolumn{3}{c|}{InAs} &
      \multicolumn{3}{c}{Ge} \\


		Density & $A$ & $B$ & $C$ & $A$ &$B$ & $C$ & $A$ & $B$ &
			$C$\\[0.08cm] \hline
			
		0.5 & 2.99 &0.27 &0.53 &9.02 &1.94 &3.60 &5.84&1.13 &1.60 \\
			
		1.0 & 2.92 &0.25 &0.58 & & & & & & \\
			
		1.5 & 2.86 &0.23 &0.61 & & & & & & \\
			
		2.0 & 2.80 &0.22 &0.64 & & & & & & \\
			
		2.5 & 2.74 &0.20 &0.67 & & & & & & \\
			
		3.0 & & & & & & & & & \\

    \end{tabular}
  \end{ruledtabular}
\end{table*}

\begin{table*}
  \caption{\label{tab:EnergyLevelSpacingsDiffMaterials} The energy
    spacing (in meV) between the HH1 and LH1 levels, as well as the
    Fermi energy for GaAs, InAs, and Ge hole inversion layers at
    various densities (in $10^{15}$~m$^{-2}$) and $N_D-N_A =
    3\times10^{20}$~m$^{-3}$. In all cases considered below
    LH1 is far above the Fermi energy, thus validating the assumption that only the
    HH1 band is occupied.}
  \begin{ruledtabular}
    \begin{tabular}{c|ccc|ccc|ccc}

      & \multicolumn{3}{c|}{GaAs} & \multicolumn{3}{c|}{InAs} &
      \multicolumn{3}{c}{Ge} \\


      Density & $E_{H1}$ & $\Delta_{11}^{HL}$ & $E_{F1}$ & $E_{H1}$
      & $\Delta_{11}^{HL}$ & $E_{F1}$ & $E_{H1}$ &
      $\Delta_{11}^{HL}$ & $E_{F1}$\\[0.08cm] \hline

      0.5 & 16.12 & 8.32 & 0.91 & 12.87 & 10.87 & 2.27 &14.72 &7.74
      & 1.63 \\

      1.0 & 19.24 & 8.73 & 1.67 & 16.38 & 11.42 & 3.47 & 18.04 &
      8.16 & 2.72 \\

      1.5 & 22.12 & 9.04 & 2.30 & 19.54 & 11.79 & 4.33 & 21.08 &
      8.46 & 3.50 \\

      2.0 & 24.83 & 9.27 & 2.83 & 22.47 & 12.06 & 5.12 & 23.91 &
      8.69 & 4.18 \\

      2.5 & 27.40 & 9.46 & 3.26 & 25.21 & 12.28 & 5.88 & 26.58 &
      8.87 & 4.80\\

      3.0 & 29.85 & 9.62 &3.64 & 27.82 & 12.45 & 6.61 & 29.13 & 9.02
      & 5.43\\

    \end{tabular}
  \end{ruledtabular}
\end{table*}

\begin{table*}
  \caption{\label{tab:EnergyLevelSpacingsInSb} The energy spacing
    (in meV) between the HH1 and LH1 levels, as well as the Fermi
    energy $E_{F1}$ for an InSb hole inversion layer with $N_{D}-N_{A}
    = 3\times10^{20}$~m$^{-3}$. At the densities below (in $10^{15}$
    m$^{-2}$), only the HH1+ and HH1$-$ subbands are occupied. Here the
    material- and structure-dependent coefficients $A$, $B$, and $C$
    are listed for $p = 0.33\times10^{15}$~m$^{-2}$. $A$ is given in multiples of
    $10^{-16}$~meV$\,$m$^{2}$, $B$ in multiples of $10^{-32}$~meV$\,$m$^{4}$,
    and $C$ in multiples of $10^{-24}$~meV$\,$m$^{3}$.}
  \begin{ruledtabular}
    \begin{tabular}{c|cccccc}
      & \multicolumn{6}{c}{InSb} \\
      Density & $E_{H1}$ & $\Delta_{11}^{HL}$ & $E_{F1}$ & $A$ & $B$ & $C$
      \\[0.08cm] \hline
	0.33 &9.69  &11.72 &2.26 &15.76 &5.84 &10.06\\
	0.50 &10.87 &12.00 &2.87 & & &\\
	0.67 &12.02 &12.23 &3.33 & & &\\
	0.83 &13.05 &12.42 &3.74 & & &\\
	1.00 &14.11 &12.59 &4.15 & & &\\
    \end{tabular}
  \end{ruledtabular}
\end{table*}

\subsection{Silicon Inversion Layers}
\label{subsec:Si}

In Si, the hole density is usually of the order of $10^{16}$
m$^{-2}$, so that $k_{F}\sim5\times10^{-8}$
m$^{-1}$. \cite{Fischetti2003} In this regime, the term $Ck^3$ in
Eq.\ (\ref{eq:SW_dispersion_with_Rashba}) becomes dominant, SW
breaks down and only numerical diagonalization gives reliable
results. As pointed out above, due to the large cubic terms, it
is necessary to include the spin-orbit split-off band, hence the
Luttinger Hamiltonian is now projected onto a $12\times12$
subspace. In this work, we consider densities up to
$2\times10^{16}$~m$^{-2}$ (Table~\ref{tab:SiEnergyLevelSpacings}),
since for larger densities the higher subband will start to
populate, violating our initial assumption.

\begin{figure*}
  \includegraphics[scale=0.85]{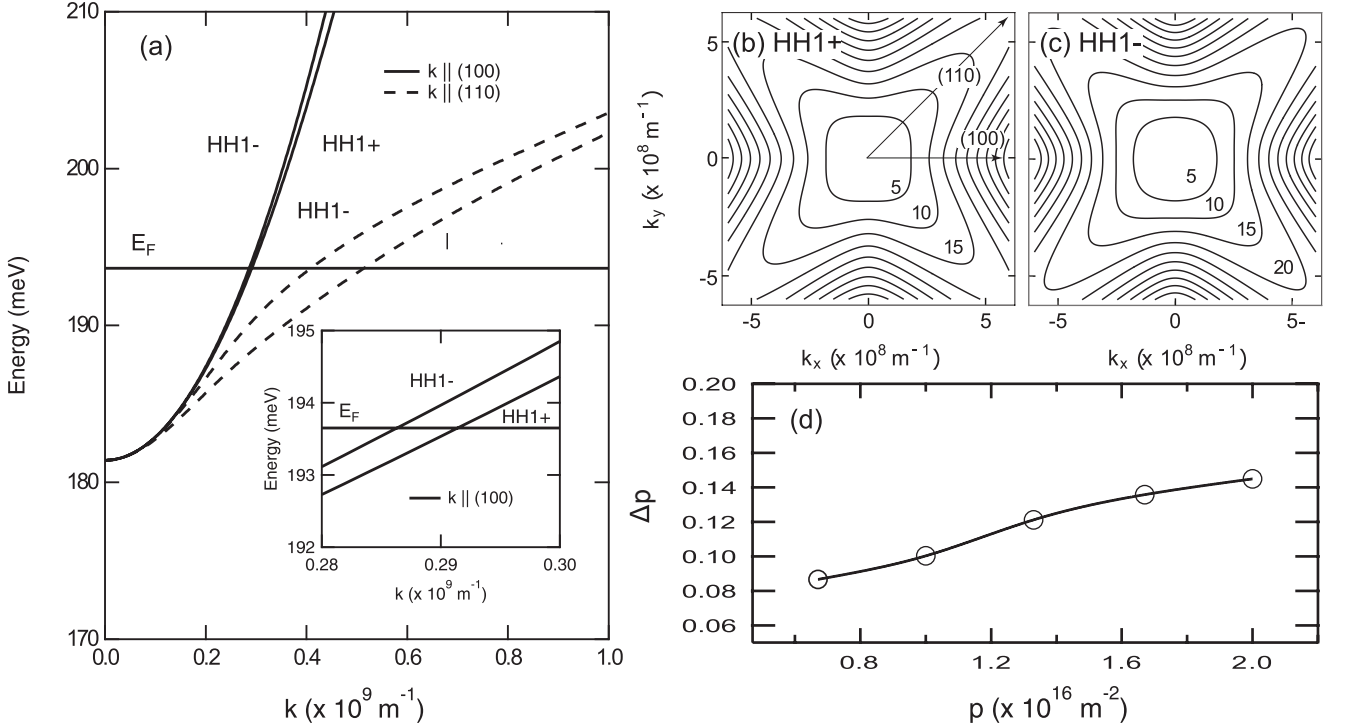}
  \caption{Panel (a) shows the dispersion for Si holes along (100)
    and (110), indicating that the strength of the spin-orbit
    interaction is a maximum along (110) and minimal along (100). The
    inset shows that the spin splitting is only $6\times{10}^{6}$
    m$^{-2}$ along $(100)$, compared to $1\times{10}^{8}$~m$^{-2}$
    along $(110)$. Here, the density is $2\times10^{16}$~m$^{-2}$
    and the doping concentration is $N_{D}-N_{A} = 3\times10^{23}$
    m$^{-3}$. The equienergy contours (separated by 5 meV) for the
    (b) HH1+ and (c) HH1$-$ subbands in the inversion layer
    discussed in panel (a) show a strong valence band warping. Panel
    (d) shows the spin splitting for Si hole inversion layers at
    various densities with $N_{D}-N_{A} =
    3\times10^{23}$~m$^{-3}$. Here, the dispersion is obtained by
    numerically diagonalizing Eq. (\ref{eq:subband_HLutt}).}
  \label{fig:Si_dispersion}
\end{figure*}

Due to the significant size of the prefactor $\zeta$ in
Eq.\ (\ref{eq:HLutt}) the SO interaction and hence spin splitting
(\ref{eq:p_plus_minus_general}) is anisotropic: It has a minimum
along the $(100)$ direction and a maximum along $(110)$. The
anisotropy of the SO strength is shown in
Fig.~\ref{fig:Si_dispersion}(a): The difference in the Fermi wave
vectors $k_{F+}$ and $k_{F-}$ is $6\times{10}^{6}$~m$^{-1}$ along
$(100)$ and $1\times{10}^{8}$~m$^{-1}$ along $(110)$. The equienergy
lines for Si ground state HHs [Figs.~\ref{fig:Si_dispersion}(b) and
  \ref{fig:Si_dispersion}(c)] show that Si valence bands are in
general warped, and the warping becomes more pronounced as the
density increases.

The spin splitting in Si [Fig.~\ref{fig:Si_dispersion}(d)] also
increases with density, which again reflects the fact that the
strength of spin-orbit interaction increases with density. However,
spin splittings in Si holes are smaller than in both zincblende materials
and Ge, discussed in the previous section. This is expected since
amongst all the materials considered in this paper, Si is the lightest
element with atomic number $Z=14$. Nevertheless, the spin splitting
in Si hole systems is considerably larger than in Si electron systems.

\begin{table}
  \caption{\label{tab:SiEnergyLevelSpacings} Si hole inversion
    layer energy level spacings (in meV) for various densities (in
    $10^{16}$~m$^{-2}$), showing that only the HH1+ and HH1$-$ bands
    are occupied. The dopant concentration $N_{D}-N_{A}$ for the
    inversion layer is $3\times10^{23}$~m$^{-3}$.}
  \begin{ruledtabular}
    \begin{tabular}{c|ccc}
      & \multicolumn{3}{c}{Si} \\ Density & $E_{H1}$ &
      $\Delta_{11}^{HL}$ & $E_{F1}$ \\ \hline 0.83 & 152.81 & 15.36
      &6.62 \\ 1.00 & 157.13 & 15.51 & 7.64 \\ 1.33 & 165.36 & 15.75
      & 9.36 \\ 1.67 & 173.63 & 15.96 &10.87\\ 2.00 & 181.48 & 16.12
      & 12.17 \\
    \end{tabular}
  \end{ruledtabular}
\end{table}

\section{Dresselhaus Spin-Orbit Interaction}
\label{sec:Dresselhaus}

In semiconductor crystals which lack a center of inversion, such as
GaAs, InAs, and InSb, Dresselhaus spin-orbit interaction terms are
present. \cite{Dresselhaus1955} In this section, we discuss the spin
splitting as a function of density and material and material
parameters in GaAs, InAs, and InSb accounting for both the Rashba
and the Dresselhaus interactions. The Dresselhaus spin-orbit
coupling in bulk semiconductor hole systems is characterized by a
variety of terms, whose relative importance depends on the parameter
regime under study. In what follows we establish a hierarchy among
these terms in heterojunctions at realistic experimental densities.

Up to third order in $k$, the Dresselhaus SO interaction in bulk
hole systems is described by the following invariants
\cite{Dresselhaus1955, Winkler2003}
\begin{equation}
  \renewcommand{\arraystretch}{1.4}
  \begin{array}[b]{rcl}
    H_{D} &= &
  -  \frac{2}{\sqrt{3}} C_D \left[k_x \{J_x ,
      J_y^2-J_z^2\}+\cp\right] \\ & &
    - B_{D1} \left[k_x\left(k_y^2-k_z^2\right) J_x +\cp\right] \\ & &
    - B_{D2} \left[k_x\left(k_y^2-k_z^2\right) J_x^3 +\cp\right] \\ & &
    - B_{D3} \left[k_x\left(k_y^2+k_z^2\right) \{J_x , J_y^2-J_z^2\}
      +\cp\right] \\ & &
    - B_{D4} \left[k_x^3 \{J_x , J_y^2-J_z^2\} +\cp\right],
  \end{array}
  \label{eq:Bulk_Dreselhaus}
\end{equation}
where $J_i, i =x,y,z$ represent the spin-3/2 matrices, and $\cp$
denotes cyclic permutation. We have one invariant linear in
$k$, quantified by the coefficient $C_D$, while the remaining four
invariants are cubic in $k$. Note that the coefficient $C_D$
is to be distinguished from the
coefficient $C$ resulting from the SW transformation. The values of
$C_D$, $B_{D1}$, $B_{D2}$, $B_{D3}$ used in this work are given in
Table~\ref{tab:LuttingerParameters}. The Dresselhaus spin-orbit coupling of bulk hole systems
typically reflects the competition between the terms with prefactors
$C_D$ and $B_{D1}$, while the contributions due to the remaining
terms are approximately two orders of magnitude smaller in magnitude
(see Table~\ref{tab:LuttingerParameters}) and can
often be neglected. \cite{Winkler2003}

According to the theory of invariants, \cite{Bir74} the effective
$2 \times 2$ Hamiltonian representing the Dresselhaus spin-orbit
coupling in the HH1$\pm$ subspace up to third order in $k$ takes the
form
\begin{equation}
  \renewcommand{\arraystretch}{1.4}
  \begin{array}[b]{rcl}
    H_{D,2\times2} &=& -
    \gamma_{D} \left( k_+ \sigma_-  + k_- \sigma_+ \right) \\ & &
    - \beta_{D1} \left( k_+ k_- k_+ \sigma_- + k_- k_+ k_- \sigma_+ \right) \\ & &
    - \beta_{D2} \left( k_+^3 \sigma_+ + k_-^3 \sigma_- \right).
  \end{array}
  \label{eq:Dresselhaus_2x2}
\end{equation}
To evaluate the prefactors in Eq.\ (\ref{eq:Dresselhaus_2x2}), we
first project Eq.\ (\ref{eq:Bulk_Dreselhaus}) onto the subspace
spanned by the zero- and one-node hole states. Then, by means of the
SW transformation, we obtain an effective $2 \times 2$ Hamiltonian
describing the Dresselhaus interactions in the HH1$\pm$ subspace, so
that the prefactors can be obtained by comparison with Eq.\
(\ref{eq:Dresselhaus_2x2}). Analytical expressions for the prefactors $\gamma_{D}$, $\beta_{D1}$, and $\beta_{D2}$ are listed and discussed in the Appendix, complemented by typical numerical values.

Strictly speaking, our treatment of Dresselhaus SO coupling relies implicitly on a hierarchy of steps. The first step, carried out in detail in Ref.~\onlinecite{Winkler2003}, is the projection of the bulk $14 \times 14$ Hamiltonian yielding the bulk $4\times 4$ Luttinger Hamiltonian, which includes all the possible bulk Dresselhaus terms (\ref{eq:Bulk_Dreselhaus}) with the prefactors listed in Table~\ref{tab:LuttingerParameters}. In the second step the bulk Luttinger Hamiltonian is projected onto the subspace spanned by the (spin-split) ground state HH1 sub-band. In this sense our approach is comparable to that of Ref.~\onlinecite{Durnev2014} which directly projects from the bulk $14 \times 14$ Hamiltonian to the HH1 subspace.

\begin{figure}
  \includegraphics[scale=0.75]{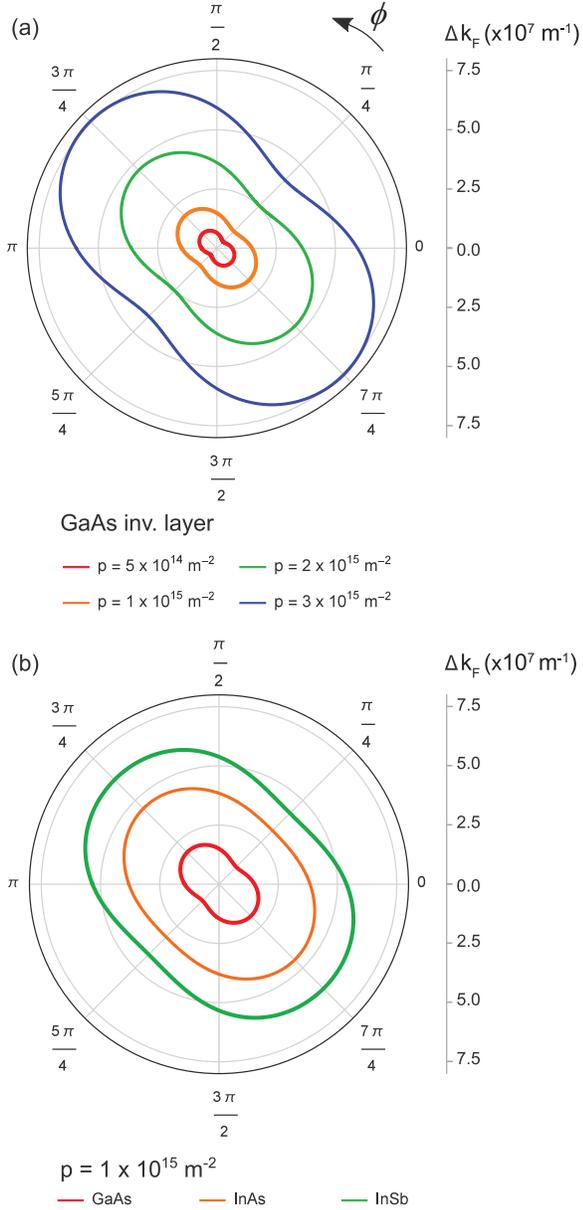}
  \caption{(Color online) Polar plot of spin splitting
    $\Delta{k_F}\equiv|k_{F+} - k_{F-}|$, which is anisotropic in
    $\bm{k}$, for (a) GaAs inversion layers at various densities and
    (b) GaAs, InAs, InSb inversion layers at $p =
    1\times{10}^{15}$~m$^{-2}$ in the presence of both Rashba and
    Dresselhaus interactions. Here, the dopant concentration is
    $N_{D}-N_{A} = 3\times10^{20}$~m$^{-3}$. The angle $\phi$ is
    defined as $\phi=\arctan{(k_y/k_x)}$.  The results here are
    obtained from a numerical diagonalization of
    Eq. (\ref{eq:subband_HLutt})}
  \label{fig:Dresselhaus_polar_plot}
\end{figure}

When only the dominant Rashba and Dresselhaus terms are retained the
energy dispersion relation becomes
\begin{widetext}
\begin{equation}
  \label{eq:SW_disp_with_Dresselhaus}
  E(k) = A k^2 - Bk^4 + 2 dk^4 (1 - \cos{4\phi}) \pm \sqrt{(|\alpha_{R2}|^2+\beta_{D1}^2)k^6 +
    2\beta_{D1}\gamma_{D}k^{4}+ \gamma_D^2k^{2} - 2
    |\alpha_R^{(2)}|(k^2\beta_{D1}+\gamma_D)k^{4}\sin{2\phi}}
\end{equation}
\end{widetext}
where $\phi = \arctan{(k_y/k_x)}$. We find that, for the single
heterojunctions studied here, the Dresselhaus terms contribute at
most one order of magnitude less than the Rashba term to the total
spin splitting at the Fermi energy. Nevertheless, the contribution
due to the Dresselhaus coupling is visible in the term proportional
to $\sin{2\phi}$, which causes the difference
$\Delta{k_F}\equiv|k_{F+} - k_{F-}|$ to be anisotropic in
$\bm{k}$. The anisotropic terms in Eq.\
(\ref{eq:2x2_Hamiltonian_with_Rashba}) are likewise sizable. In
accordance to Eq.\ (\ref{eq:Dresselhaus_2x2}), the extrema of
$\Delta{k_F}$ occur when $\phi = \pi/4$ or
$\phi = 3\pi/4$. The anisotropy of the spin splitting, which
we define here as the ratio
$\kappa \equiv \Delta{k_F} (\phi=3\pi/4) / \Delta{k_F}
(\phi=\pi/4)$,
depends on the density as well as on material-specific parameters,
as Fig.~\ref{fig:Dresselhaus_polar_plot} shows.  For example, one
can infer from Fig.~\ref{fig:Dresselhaus_polar_plot}(a) that
$\kappa = 2.24$ for a GaAs inversion layer with
$p = 5\times{10}^{14}$ m$^{-2}$ and $\kappa = 1.77$ with
$p = 3\times{10}^{15}$ m$^{-2}$. The fact that the anisotropy
$\kappa$ decreases with density implies that the Rashba coefficient
increases faster with density than the cubic and linear Dresselhaus
coefficients combined (see Appendix).  Comparing different materials
[Fig.~\ref{fig:Dresselhaus_polar_plot}(b)], one can deduce that
$\kappa = 1.89$, $\kappa = 1.42$, $\kappa = 1.48$ for GaAs, InAs,
and InSb inversion layers with $p = 1\times{10}^{15}$~m$^{-2}$,
respectively. This implies that, amongst the materials considered
here, the effect of Dresselhaus SOI is weakest in InAs but strongest
in GaAs, an observation that is rather counterintuitive considering
the relative values of the bulk Dresselhaus prefactors in these
materials.  Indeed, this reflects the fact that the relative importance of specific spin-orbit interaction terms is determined by their ratio to the spin-independent terms in the dispersion relation, rather than the absolute magnitude of their corresponding numerical prefactors.

\section{Discussion}
\label{sec:Discussion}

\subsection{Breakdown of the SW method}

The analysis presented in this work is consistent with the known
general trend that SO coupling in 2DHGs increases with the carrier
number density. We have characterized the strength of the Rashba SO
interaction using both the relative spin-split population difference
$\Delta p$ and the effective masses $m_{\pm}$. We have found that
$\Delta p$ and $m_+$ are expected to increase almost linearly with
density whereas $m_-$ is approximately constant. While the simple
dispersion found using the SW transformation provides a good
approximation for certain systems, e.g. GaAs inversion layers with
$N_{D}-N_{A} = 3\times10^{20}$~m$^{-3}$ and densities $p <
2.5\times10^{15}$~m$^{-2}$, there are many parameter regimes in
which SW breaks down. The fact that the simple dispersion
can fail for realistic parameters challenges the current
theories on electric dipole spin resonance (EDSR), \cite{Bulaev2007}
spin conductivity, \cite{Schliemann2004, Sugimoto2006, Wong2010,
  Bi2013} hole spin helix, \cite{Sacksteder2014} and
\textit{Zitterbewegung}, \cite{Schliemann2006, Biswas2015} which
assume that hole spin-orbit coupling is simply cubic in wave vector.

\subsection{Inversion vs. accumulation layers}

Comparing GaAs inversion and accumulation layers, the strength of
the Rashba SO interaction is very sensitive to the type of
background dopant. Owing to the fact that the confinement potential
for inversion layers is steeper than that for accumulation layers,
the HH1-LH1 splitting for inversion layers is larger than that for
accumulation layers. Consequently, the strength of SO interaction is
stronger in an accumulation layer than in an inversion layer (see
Appendix for the relation between Rashba or Dresselhaus coefficients
and the HH1-LH1 splitting).

In both inversion and accumulation layers, the spin splitting
$\Delta p$ increases with density, yet the dependence of the spin
splitting and effective mass on density is quite different. In an
accumulation layer, both $\Delta p$ and $m_+$ increase almost
linearly with $p$ at low densities, and both saturate at higher
densities. These trends can be attributed to the fact that in an
accumulation layer, the HH1-LH1 anticrossing occurs at a lower $k$,
so that the spin splitting near the anticrossing is
reduced. Correspondingly, in accumulation layers SW breaks down at
much lower densities than in inversion layers.

We compare our results for GaAs, InAs, InSb, Ge and Si inversion
layers. The spin splitting and effective mass profiles show that the
Rashba SO interaction is stronger in Ge, InAs, and InSb than in
GaAs, though the Fermi contour remains isotropic in the absence of
Dresselhaus terms. The Si dispersion, on the other hand, is very
anisotropic owing to the large cubic contribution stemming from the
bulk Luttinger Hamiltonian.\cite{Winkler2003} The term $\propto
\zeta$ in the Luttinger Hamiltonian Eq.~(\ref{eq:HLutt}) is sizable
and is responsible for the highly anisotropic Fermi contour. The
spin splitting is a maximum along the $(110)$ and $(1\bar{1}0)$
directions but a minimum along $(100)$ and $(010)$.

We have considered the Dresselhaus spin-orbit interaction in
zincblende semiconductors GaAs, InAs, and InSb. While the magnitude
of BIA is always much smaller than SIA, it causes the spin splitting
to be anisotropic in $\bm{k}$. Figure \ref{fig:Dresselhaus_polar_plot}
shows that the spin splitting is a maximum along $(110)$ but a
minimum along $(1\bar{1}0)$, in accordance with
Eq.\ (\ref{eq:SW_disp_with_Dresselhaus}).

\subsection{Relative importance of the $k$-linear and $k$-cubic spin-orbit interaction terms}
\label{subsec:LinearAndCubicTerms}

In general, spin-orbit interaction terms in the system Hamiltonian
are characterized by invariants, with each invariant associated with
a specific power of $k$ and having a particular functional form. The
fact that spin-orbit interactions preserve time reversal symmetry
\cite{Winkler2003} implies that only odd powers of the wave vector
are admissible. To date spin-orbit interaction terms of orders
linear\cite{Rashba1988, Bi2013, Winkler2003, Luo2010, Durnev2014,
  Wenk2016} and cubic in $k$ \cite{Dresselhaus1955, Winkler2000,
  Winkler2002, Winkler2003, Bulaev2005, Sugimoto2006,
  Schliemann2006, Winkler2008, Wong2010, Bi2013, Sacksteder2014,
  Biswas2015, Wenk2016} have received the most attention. The
identification of the relevant powers of the wave vector
characterizing a specific form of the spin-orbit interaction depends
on the underlying mechanism (BIA, SIA, and interface asymmetry), on the dimensionality of
the system, and on whether the carriers are electrons or holes. For
example, in bulk electron systems the leading-order Dresselhaus term
is $\propto k^3$, while in bulk hole systems as well as in 2D
electron and hole systems Dresselhaus terms linear in $k$ are
present. Broadly speaking, one expects terms involving smaller
powers of $k$ to dominate at low densities, while terms involving
higher powers of $k$ are expected to dominate at higher densities.

In the structures we have studied, although $k$-linear Rashba terms
of the form ${i}\alpha_{R1} (k_+\sigma_+ - k_-\sigma_-) \equiv - 2
\alpha_{R1} (k_y \sigma_x + k_x \sigma_y)$ are allowed by
symmetry,\cite{Winkler2003} we find the $k$-linear Rashba
coefficient to be zero since our model does not include excited bulk
conduction bands, as discussed in
Sec. \ref{subsec:SW}. Correspondingly, $k$-linear Rashba terms have
not been included in our analytical (SW) results. Thus, when the SW
transformation is applicable, the Rashba spin splitting is described
by Eq.\ (\ref{eq:SW_dispersion_with_Rashba}), which is consistent
with Ref.\ \citenum{Winkler2003}. In contrast, the Dresselhaus
spin-orbit interaction is characterized by terms proportional to
both $k$ and $k^3$ which are comparable in magnitude. Therefore, we
have included both $k$-linear and $k$-cubed terms in
Eq.\ (\ref{eq:Dresselhaus_2x2}).

We would like to comment on the relative importance of the
$k$-linear \cite{Winkler2003, Luo2010, Bi2013, Wenk2016} and
$k$-cubic \cite{Dresselhaus1955, Winkler2003, Bulaev2005, Wong2010,
  Bi2013, Biswas2015, Wenk2016} Dresselhaus terms for the systems
studied in this work. For GaAs inversion layers with the 2DHG
densities considered here, the linear-$k$ and cubic-$k$ Dresselhaus
terms are of the same order of magnitude (see Appendix). For
example, for a GaAs inversion layer with $p =
5\times10^{14}$~m$^{-2}$ and $k_F = 5.6\times10^{7}$~m$^{-1}$, we
find $\beta_{D1} = 0.080\times10^{-24}$ meV m$^3$, and $\gamma_D =
0.030\times10^{-8}$ meV m, so that the ratio $\xi_D \equiv
\frac{\gamma_D k_F}{\beta_{D1} k_F^3}$ of the $k$-linear and $k$-cubic
Dresselhaus terms is $\xi_D = 1.19$. As the density increases, the
relative importance of the linear-$k$ term decreases. For example,
for $p = 2\times10^{15}$~m$^{-2}$ corresponding to $k_F \approx
1.12\times10^{8}$~m$^{-1}$, we find $\beta_{D1} = 0.084\times10^{-24}$
meV m$^3$, and $\gamma_D = 0.029\times10^{-8}$ meV m, yielding
$\xi_D = 0.32$, which is four times less than when $p =
5\times10^{14}$~m$^{-2}$. Since $\beta_{D1}$ and $\gamma_D$ are almost
constant at the densities considered in this work, we can easily
estimate the density $p_{D}$ above which the cubic-$k$ Dresselhaus
term dominates. Taking $\beta_{D1} = 0.08\times10^{-24}$ meV m$^3$ and
$\gamma_D \approx 0.03\times10^{-8}$ meV m, we find that $\xi_D \ll
1$ when $k_F \gg 6.1 \times 10^7$ m$^{-1}$, which corresponds to $p
\gg 6 \times 10^{14}$~m$^{-2}$.

We have performed a similar analysis of the $k$-linear and $k$-cubic
Dresselhaus terms for InAs and InSb inversion layers. For an InAs
inversion layer with $p = 5\times10^{14}$~m$^{-2}$ we obtain $\xi_D
= 5.2$, which implies that the linear-$k$ term is much more
important than the cubic-$k$ Dresselhaus term for the densities
considered here. For InAs we find $p_D = 2.6 \times 10^{15}$
m$^{-2}$. On the other hand, for an InSb inversion layer with $p =
3.3\times10^{14}$~m$^{-2}$, $\xi_D = 0.16$. We obtain $p_D = 5.4
\times 10^{13}$ m$^{-2}$, so that the linear-$k$ contribution is
essentially negligible for our range of densities.

The Dresselhaus interaction is absent in semiconductors with a
diamond structure such as Si and Ge. At the same time terms of the
same symmetry may appear depending on the termination at the
interface. \cite{Durnev2014} The interface-induced spin splitting
depends on the coupling to the conduction and split-off bands. The
dominant contribution from this mechanism is linear in $k$ and has
the same symmetry as the $k$-linear Dresselhaus term, i.e.
$H_{\mathrm{int}} \propto k_- \sigma_+ + k_+ \sigma_- = k_x \sigma_x
+ k_y \sigma_y$.
\cite{Durnev2014} A quantitative description of interface asymmetry
terms is beyond the scope of the present work.  We may expect that
for the heterojunctions studied here the effect of interface
asymmetry is weaker than in quantum wells, since the wave
functions experience a strong confinement at the heterointerface, but only a weak confinement towards the substrate.  Thus
even in a refined model permitting the wave functions to tunnel into
the barrier, the wave functions are less pushed into the barrier
than in a quantum well. The smaller probability of finding the
carriers at the interface then implies a smaller effect of
interface asymmetry on spin splitting. Consistent with this qualitative reasoning, initial studies reveal that in heterojunctions the terms due to interface asymmetry are considerably smaller than the Rashba spin-orbit interaction. \cite{Durnev_private_comm_2016}

\section{Summary and Outlook}
\label{sec:Summary}

We have performed a variational analysis of spin-orbit interactions
in 2D hole gases in inversion-asymmetric heterojunctions in a number of cubic
semiconductors. We have quantified our findings in terms of
experimentally accessible quantities: carrier number density,
effective masses, and spin splitting. We find that for a broad range of
experimentally relevant parameters the frequently used lowest-order expansion of the dispersion breaks down. To address this shortcoming we have provided
a simple quasianalytical scheme for calculating spin-orbit related
quantities that is in good agreement with numerical studies and
experimental data. We recover the known general trend that the spin
splitting and HH1+ effective mass $m_+$ increase as functions of
density. We have found that in heterojunctions the Rashba SO coupling is
in general much stronger than the Dresselhaus SO coupling.  More specifically, 
Rashba SO coupling is much stronger in accumulation layers than in inversion layers, and it is very sensitive to the density of background
dopants. Finally, in Si, due to the strong cubic terms already
present in the bulk, the Fermi contour is strongly anisotropic.

The approach presented in this work can be extended to study spin-orbit interactions on surfaces other than (001), while the choice of basis functions can be tailored to the system under consideration. For example, the bulk $k^3$-Dresselhaus interaction (\ref{eq:Bulk_Dreselhaus}) applied to a low-symmetry surface includes terms $\propto k_z^3$, which can be problematic depending on one's choice of basis functions. Reference~\onlinecite{Durnev2014} thus avoids dealing directly with the Dresselhaus terms (\ref{eq:Bulk_Dreselhaus}) in determining the spin splitting for 2D hole systems.  Instead, this work resorts to the bulk $14 \times 14$ extended Kane Hamiltonian, which avoids powers of $k_z$ higher than second, offering considerable computational flexibility. Nevertheless, although higher powers of $k_z$ require some care, they can easily be treated if the basis functions are sufficiently smooth (e.g., plane waves \cite{WinklerRoessler1993} or Fock-Darwin states). This consideration can become important if in certain systems terms $\propto k_z^3$ turn out to be significant, which is not the case in the present work. 

\section*{Acknowledgements}

This work has been supported by the Australian Research Council through the Discovery Project scheme. R.W.
was supported by the NSF under Grant No.\ DMR-1310199.  Work at
Argonne was supported by DOE BES under Contract
No.\ DE-AC02-06CH11357. We thank Ulrich Zuelicke, Daisy Wang, Oleg Sushkov,
Tommy Li, and Dima Miserev for insightful comments and
discussions. We are also indebted to Scott Liles and Ashwin
Srinivasan for providing their experimental data for comparison.

\appendix*

\section{The material- and structure-dependent parameters}
\label{sec:ABC_Coeffs}

In the following we evaluate the prefactors up to second order
perturbation theory.  The first-order terms are given by the
expectation values of the diagonal elements in the $2\times2$
submatrix spanned by the states $\{\ket{\pm3/2}\}$ in the Luttinger
Hamiltonian (\ref{eq:subband_HLutt}).  The second-order terms are
mediated by couplings to intermediate states in the usual
way. \cite{Winkler2003} In all formulas in this section, the index
$i$ implies a summation over $i = 1, 2$. 

The prefactors for the orbital terms take the form
\begin{align}
  A = & \frac{\mu}{2} (\gamma_1+\gamma_2)
     + 3 \mu^2 \,\gamma_3^2 \frac{|\braket{H_1 | k_z | L_i}|^2}
     {\Delta_{1i}^{HL}} \\
  B = & -\frac{3}{4} \, \mu^2 (\bar{\gamma} - \zeta)^{2}
  \frac{{|\braket{H_1 | L_i}|^2}}{\Delta_{1i}^{HL}} \\
  d = & \frac{3}{4} \,\bar{\gamma} \, \zeta
  \frac{{|\braket{H_1 | L_i}|^2}}{\Delta_{1i}^{HL}}
\end{align} 
where $\Delta_{pq}^{rs} \equiv E_p^r-E_q^s $.  Typical values for
$A$, $B$ and $d$ for GaAs inversion layers are given in Table
\ref{tab:A_and_B_second_order}.

\begin{table*}
 \caption{\label{tab:A_and_B_second_order} Typical values for
   $A$ (in $10^{-16}$~meV$\,$m$^{2}$),
   $B$, $d$ (both in $10^{-32}$~meV$\,$m$^{4}$),
   $\alpha_{R2}$, $\alpha_{R3}$
   (both in $10^{-24}$~meV$\,$m$^{3}$),
   $\gamma_D$ (in $10^{-8}$~meV$\,$m),
   $\beta_{D1}$ and $\beta_{D2}$ (both in $10^{-24}$~meV$\,$m$^{3}$),
   for 2DHG GaAs hole inversion layers with
   $N_{D}-N_{A} = 3\times10^{20}$~m$^{-3}$.  Superscripts $(1)$
   and $(2)$ distinguish contributions from first- and second-order
   perturbation theory to the respective coefficients. The density is given in
   multiples of $10^{15}$~m$^{-2}$.}

 \begin{ruledtabular}
   \begin{tabular}{c*{5}{d{1,2}}*{5}{d{1,3}}}
     \multicolumn{1}{c}{density} & \mc{$A^{(1)}$} & \mc{$A^{(2)}$} &
     \mc{$B$} & \mc{$d$} &
     \mc{$\alpha_{R2}$} & \mc{$\alpha_{R3}$} &
     \mc{$\gamma_D^{(1)}$} & \mc{$\gamma_D^{(2)}$} & 
     \mc{$\beta_{D1}$} & \mc{$\beta_{D2}$} \\ \hline
     0.5 & 3.41 & -0.42 & 0.27 & -0.70 & -0.53 & 0.084 & 0.022 & 0.008 & 0.083 & -0.013 \\
     2.0 & 3.41 & -0.61 & 0.22 & -0.56 & -0.64 & 0.103 & 0.017 & 0.010 & 0.086 & -0.014 \\
   \end{tabular}
 \end{ruledtabular}
\end{table*}

The Rashba coefficient $\alpha_{R2}$ reads
\begin{equation}
  \label{eq:Rashba_second_R2}
  \alpha_{R2} = -\frac{3}{2} \, \mu^2 \gamma_{3}
  \bar{\gamma}
  \frac{\braket{H_1 | L_i} \braket{L_i | k_z | H_1}
  - \braket{H_1 | k_z | L_i} \braket{L_i | H_1}}{\Delta_{1i}^{HL}},
\end{equation}
while the Rashba coefficient $\alpha_{R3}$ is given by 
\begin{equation}
  \label{eq:Rashba_second_R3}
  \alpha_{R3} = \frac{3}{2} \, \mu^2 \gamma_{3}
  \zeta
  \frac{\braket{H_1 | L_i} \braket{L_i | k_z | H_1}
  - \braket{H_1 | k_z | L_i} \braket{L_i | H_1}}{\Delta_{1i}^{HL}}.
\end{equation}
implying $\alpha_{R2} / \alpha_{R3} = - \bar{\gamma} / \zeta$.
Note that these expressions emerge in second order in the SW
transformation.  This is in contrast to
Ref.~\onlinecite{Winkler2003}, where these prefactors arose only in
third order.  This difference is due to the fact that in
Ref.~\onlinecite{Winkler2003}, the prefactors were expressed in
terms of basis functions for inversion symmetric systems such as the
infinite square well and the simple harmonic oscillator, treating
the asymmetric component of the confining potential $V(z)$
explicitly as a perturbation.  In this work, the asymmetry of $V(z)$
is encoded in the asymmetric Fang-Howard functions
(\ref{eq:FHGroundState-1}) and (\ref{eq:FHExState}), so that the
expressions (\ref{eq:Rashba_second_R2}) and
(\ref{eq:Rashba_second_R3}) for Rashba prefactors depend implicitly
on the asymmetry of $V(z)$.

The $k$-linear Dresselhaus coefficient has the form:
\begin{equation}
\label{eq:Dresselhaus_linear_second}
  \renewcommand{\arraystretch}{1.4}
  \begin{array}[b]{rs{0.2em}L}
  \gamma_D  = & -\frac{\sqrt{3}}{2}C_D
  +2\sqrt{3} \mu \, \gamma_3 C_D
	  \frac{|\braket{H_1 | k_z | L_i}|^2}{\Delta^{HL}_{1i}} \\
 &  - \frac{3}{4}(B_{D2}+B_{D3})\braket{H_1|k_z^2|H_1}
\end{array}
\end{equation}
The prefactors of the cubic Dresselhaus terms become
\begin{equation}
  \label{eq:Dresselhaus_cubic_D1}
  \renewcommand{\arraystretch}{1.4}
  \begin{array}[b]{L}
    \beta_{D1} = \frac{\sqrt{3}}{2} \mu \, \bar{\gamma} \, C_D
    \frac{|\braket{H_1 | L_i}|^2}{\Delta^{HL}_{1i}}
    - \frac{3}{16} (B_{D2} - B_{D3} + 3 B_{D4})
    \\ \hspace{1em}
    + \frac{3}{4} \mu \, \bar{\gamma} B_{D1} \,
    \frac{\braket{H_1 | L_i} \braket{L_i | k_z^2 | H_1}
    + \braket{H_1 | k_z^2 | L_i} \braket{L_i | H_1}}{\Delta_{1i}^{HL}}
    \\  \hspace{1em}
    + \mu \bar{\gamma} \left(\frack{21}{16} B_{D2} + \frack{3}{8} B_{D3} \right) \frac{\braket{H_1|k_z^2|L_i}\braket{L_i|H_1} + \braket{H_1|L_i} \braket{L_i|k_z^2|H_1}}{\Delta_{1i}^{HL}} \\  \hspace{1em}
    + 3 \mu \gamma_3 B_{D3} \frac{|\braket{H_1|k_z|L_i}|^2}{\Delta_{1i}^{HL}} 
  \end{array}
\end{equation}
and
\begin{equation}
  \label{eq:Dresselhaus_second_D2}
  \renewcommand{\arraystretch}{1.4}
  \begin{array}[b]{L}
    \beta_{D2} = - \frac{\sqrt{3}}{2} \mu \, \zeta \, C_D
    \frac{|\braket{H_1 | L_i}|^2}{\Delta^{HL}_{1i}}
    + \frac{3}{16} (B_{D2} - B_{D3} + B_{D4})    \\  \hspace{1em}
    - \frac{3}{4} \mu \zeta B_{D1} 
    \frac{ \braket{H_1 | L_i} \braket{L_i | k_z^2 | H_1} 
    + \braket{H_1 | k_z^2 | L_i} \braket{L_i | H_1}}{\Delta_{1i}^{HL}} 
    \\  \hspace{1em}
    - \mu \zeta \left(\frack{21}{16} B_{D2} + \frack{3}{8} B_{D3} \right) \frac{\braket{H_1|k_z^2|L_i}\braket{L_i|H_1} + \braket{H_1|L_i} \braket{L_i|k_z^2|H_1}}{\Delta_{1i}^{HL}} 
  \end{array}
\end{equation}
For the $k$-linear coefficient $\gamma_D$, the bulk prefactors
$B_{D2}$ and $B_{D3}$ give rise to sizable first-order
contributions at low densities.  On the other hand, interestingly,
there are no contributions proportional to the bulk prefactors
$B_{D2}$, $B_{D3}$, and $B_{D4}$ from second order in SW
perturbation theory.  The smallness of $B_{D2}$,
$B_{D3}$, and $B_{D4}$ implies approximately
$\beta_{D1} / \beta_{D2} \simeq - \bar{\gamma} / \zeta$.
Nonetheless, for the prefactors $\beta_{D1}$ and $\beta_{D2}$,
we retained the contributions up to second order in SW perturbation theory.  
Here we note that within the extended Kane
model we have the relation $B_{D4} = B_{D3} - B_{D2}$ (see
Table~\ref{tab:LuttingerParameters} and
Ref.~\onlinecite{Winkler2003}), so that the first-order
contributions from the coefficients $B_{D2}$, $B_{D3}$, and $B_{D4}$
to the coefficient $\beta_{D2}$ cancel. While the contributions from the bulk Dresselhaus terms proportional to $B_{D2}$, $B_{D3}$, and $B_{D4}$ are generally small, the terms obtained in first and second-order SW perturbation theory are similar in magnitude.  This reflects the fact that the second order yields mixed terms still linear in  $B_{D2}$ and $B_{D3}$, but also proportional to the large Luttinger coefficients.

Typical values for the Rashba and Dresselhaus coefficients obtained from the above equations are given in Table~\ref{tab:A_and_B_second_order}.


%

\end{document}